\documentclass[preprint,12pt]{elsarticle}

\usepackage{amssymb}
\usepackage{lineno,hyperref}
\usepackage{graphicx}
\usepackage{amssymb}
\usepackage{latexsym}
\usepackage{amsmath}
\usepackage{algorithm}
\usepackage{algpseudocode}
\usepackage{graphics}
\usepackage{epsfig}
\usepackage{multirow}
\usepackage{subfigure}
\usepackage{float}
\usepackage{chngpage}
\usepackage{booktabs}
\usepackage{threeparttable}
\usepackage{underscore}
\usepackage[misc]{ifsym}
\modulolinenumbers[5]

\usepackage[]{caption2}

\begin{document}

\begin{frontmatter}



\title{Leveraging tagging and rating for recommendation: RMF meets weighted diffusion on tripartite
graphs}

\author[mymainaddress]{Jianguo Li}

\author[mymainaddress]{Yong Tang}

\author[mymainaddress]{Jiemin Chen\corref{mycorrespondingauthor}}
\cortext[mycorrespondingauthor]{Corresponding author}
\ead{chenjiemin@m.scnu.edu.cn}

\address[mymainaddress]{School of Computer Science, South China Normal University, Guangzhou, China}

\begin{abstract}
Recommender systems (RSs) have been a widely exploited approach to solving the information overload problem. However, the performance is still limited due to the extreme sparsity of the rating data. With the popularity of Web 2.0, the social tagging system provides more external information to improve recommendation accuracy. Although some existing approaches combine the matrix factorization models with co-occurrence properties and context of tags, they neglect the issue of tag sparsity without the commonly associated tags problem that would also result in inaccurate recommendations. Consequently, in this paper, we propose a novel hybrid collaborative filtering model named WUDiff_RMF, which improves Regularized Matrix Factorization (RMF) model by integrating Weighted User-Diffusion-based CF algorithm(WUDiff) that obtains the information of similar users from the weighted tripartite user-item-tag graph. This model aims to capture the degree correlation of the user-item-tag tripartite network to enhance the performance of recommendation. Experiments conducted on four real-world datasets demonstrate that our approach significantly performs better than already widely used methods in the accuracy of recommendation. Moreover, results show that WUDiff_RMF can alleviate the data sparsity, especially in the circumstance that users have made few ratings and few tags.
\end{abstract}


\end{frontmatter}


\section{Introduction}
\label{intro}
The recent decade has witnessed the rapidly increasing amount of information available on the Internet, which make people engulf in a wide variety of information. To solve this problem, recommender systems (RSs) have been applied for suggesting products in different domains, such as music recommendation in Last.fm, movie recommendation at Netflix and video recommendation at YouTube. The objective of RSs is to assist users to find out suitable items based on the users' preference in the past.\par

As a prominence of recommendation techniques, collaborative filtering (CF) methods [1] recommend items based on the preference of other similar items or like-minded users, including the neighborhood method and the matrix factorization model [2]. The former intends to predict a user's rating on an item from like-minded users or similar items. By comparison, the latter uses the user-item matrix to map both users and items into a latent factor space for representing their relationship with high accuracy and scalability. Traditional CF methods generally only use the user-item rating matrix for recommendation. However, because the density of available ratings is often less than 1\%, traditional CF methods suffer from data sparsity and the cold-start problem that remarkably reduce the system performance [3,4].\par

With the dramatic development of Web 2.0, social tag systems have emerged and become popular, which allow users to freely assign, organize and share items with tags [5]. For instance, the song \lq\lq Roll in the Deep\rq\rq  are evaluated by 307,915 users and labeled with 61 tags on Last.fm, where 5091 users tag the song with \lq\lq amazing voice\rq\rq  and \lq\lq best of 2011\rq\rq. User-defined tags are used as a means to represent their preference and evaluations for the items. Therefore, based on the intuition that users' tagging history can be utilized to improve the performance of recommender systems, the research of tag-based recommendation has become popular, especially tag-based CF methods [6]. Most of them are unified frameworks based on the technique of matrix factorization, which firstly use tagging data to obtain the neighborhood information of each item(user), and then incorporated the information into a probabilistic factor analysis model [6,7]. However, in some real-world scenarios, the user rating data and tagging data are very sparse, so the neighborhood information may be inaccurate. For example, suppose Kent, Mark and Sam love comedy movies. Kent has rated three comedy movies with the high ratings and Mark has rated other five comedy movies with his high ratings. At the same time, Mark has labeled this movie \lq\lq Dead pool\rq\rq with two tags and Sam has labeled the movie with other three tags. Due to the lack of co-rated items, rating-based CF methods may ignore that Kent and Mark share common preference on comedy movies. And without co-occurrence, some tagging-based CF methods also may neglect that Mark and Sam share similar preference. The inaccurate information may result in misleading recommendations.\par

The preference of users should include tags and ratings. Consequently, inspired by the idea of network-based models from statistical physics, we propose the RMF with WUDiff (WUDiff_RMF) hybrid recommendation approach. Firstly, we represent users and items with a joint latent factor space of dimensionality by RMF. Secondly, we obtain the relationship of similar users by Weight User-Diffusion algorithm(WUDiff), which apply a diffusion process to generate recommendations in a user-item-tag tripartite graph. Finally, we expand the RMF by integrating similar user regularization term which describes that the target user would be influenced by similar users in the weighted tripartite network. A series of experiments are conducted on four popular real-world datasets to validate the effectiveness of WUDiff_RMF. The results confirm that our approach can effectively improve the accuracy of predictions with the other counterparts. And the further analysis indicates it can alleviate the data sparsity problem especially in datasets with each user making sparse ratings and tagging data.\par

The rest of the paper is organized as follows. Section 2 briefly describes related work. Section 3 describes preliminaries, revisits the RMF model and the Udiff recommendation algorithm. We describe a new, more accurate hybrid recommendation model in Section 4. The experimental results are presented and analyzed in Section 5. Finally, we conclude the paper and give future research directions in Section 6.\par

\section{Related work}
\label{Relatedwork}
Tagging systems provide users with a promising way to freely associate tags with web resources, which are good at describing users' opinions. Thus, researchers endeavor to improve recommendation accuracies by utilizing tagging data, especially for those CF methods. In this section, we review several major tag-based CF recommendation approaches, which are generally divided into the neighborhood method and the model method.\par

The measurement of similarity between users or items has played a vital role in the neighborhood-based CF. By taking the relationships among tags into consideration, the co-occurrence properties and context of tags can be employed to obtain the neighbors of each user and each item. Sen et al.[8] constructed implicit and explicit tag-based recommendation algorithms based on the user-tag rating matrix. Wang et al.[9] improved the tag-based neighborhood method by using tagging data to generate latent topics. Qi et al.[10] computed users' similarities by utilizing the inferred tag ratings for improving user-based CF method. Gedikli et al.[11] improved the item-based CF by incorporating tag preferences in the context of an item. In addition, from a perspective of graph theory, a user-item-tag-based network can be viewed as a tripartite network[5]. Zhang et al.[12] firstly proposed a tag-aware diffusion-based method(ODiff) that represents tags as nodes in a tripartite graph and utilizes a diffusion process to obtain better recommendations. Shang et al.[13] proposed an user-based hybrid tag algorithm by harnessing diffusion-based method(UDiff). Trinity is introduced in [14], where a random walk with restart model is proposed based on a three-layered object-user-tag heterogeneous network.\par

As one of the most successful and popular model-based CF methods, Matrix factorization (MF) includes Regularized Matrix Factorization (RMF) [15], Probabilistic Matrix Factorization (PMF) [16], principle component analysis (PCA)[17], latent Dirichlet allocation (LDA) [18] and so on. MF can capture overall structure that associates with most or all items, but overlooks strong associations among a small set of closely related neighbor set [6]. Recently, a trend in the literature is the use unified frameworks that combine neighbor-based methods and MF models to handle the data sparsity problem [7,19]. Zhou et al.[20] proposed a factor analysis approach called TagRec based on a unified probabilistic matrix factorization by utilizing both users¡¯ tagging information and rating information. Wu et al.[6] built a two-stage recommendation framework, named NHPMF, by using the tagging data to select neighbors of each user and each item. Zhang et al.[21] proposed a recommendation model based on clustering of users (UCMF) by considering the neighbors' impact on the interest of each user in the same latent factor space. Chen et al.[22] proposed TRCF model that improved recommendation performance by capturing the semantic correlation between users and items. Additionally, from the view of feature level, Zhang et al.[23] presented the feature-centric recommendation approach that utilized user's feature preferences to improve recommendation of items.\par

Existing approaches incorporates localized relationships into MF model by using co-occurrence properties and context of tags for finding the set of neighbors. If the user only associates very few tags with items, these approaches cannot obtain better performance. The feedback information of users should include tags and ratings. Therefore, we propose the WUDiff-based RMF model(WUDiff_RMF), which is a novel unified recommendation framework based on tags and ratings. Different from previous work, based on the information of similar users from the weighted tripartite user-item-tag network, we integrate similar user regularization term into RMF for improving the performance of recommendation.\par

\section{Preliminaries}
\label{Preliminaries}
In a standard setting of CF, we use $ U=\{{{u}_{1}},{{u}_{2}},...,{{u}_{m}}\} $ to denote the set of users, the set of items is $I=\{{{i}_{1}},{{i}_{2}},...,{{i}_{n}}\}$ and the set of tags is $T=\{{{t}_{1}},{{t}_{2}},...,{{t}_{k}}\}$. $R\left| U \right| \times \left| I \right|$ is the user-item rating matrix and $R'\left| U \right| \times \left| T \right|$ is the user-tag tagging matrix. ${r_{ui}}$ is the rating by user u of item i, which can be binary or integers from a given range (e.g. ${r_{ui}} \in [1,5]$). Notations used in this paper are summarized in Table 1.\par
%
\begin{table}[H]
\caption{ List of symbols}
\label{tab:1}       
\begin{tabular}{ll}
\hline\noalign{\smallskip}
symbol &meaning \\
\noalign{\smallskip}\hline\noalign{\smallskip}
$\kappa$&a set of ${r_{ui}}$ that are known \\
$f$& the number of latent factor \\
$\widehat {{r_{ui}}}$ & the predicted value of ${r_{ui}}$ \\
${p_u},{q_i}$ & $f$ dimension vectors of user u and item $i$\\
$R(u)$ & the set of items that are rated by user u\\
$S(u)$ & the set of users who are similar to user u by WUDff\\
$A$ & user-item adjacent matrix\\
${\alpha _{{\rm{ij}}}}$ & an edge between user $i$ and item $j$\\
$A'$ & user-tag adjacent matrix\\
${\alpha '_{ik}}$ & an edge between user $i$ and tag $k$\\
$k(user{}_i)$ & the degree of user $i$\\
$k(item{}_j)$ & the degree of item $j$\\
$k(tag{}_k)$ & the degree of tag $k$\\
\noalign{\smallskip}\hline
\end{tabular}
\end{table}

\subsection{Regularized Matrix Factorization}
\label{Regularized Matrix Factorization}
RMF is an efficient and effective approach to RSs among a large number of solutions to RSs [24]. The fundamental principle of RMF maps both users and items to a low-dimensional feature space of dimensionality $f$, and utilizes the latent factor space to make further missing data prediction.  Accordingly, $P \in {R^{\left| U \right| \times f}}$ denotes the user feature matrix and each user $u$ is associated with a latent fact vector ${p_u}$ that represents users personal interests. $Q \in {R^{\left| I \right| \times f}}$ represents the item feature matrix. Each item $i$ is associated with a latent fact vector ${q_i}$ to describe the characteristics. Then the rating approximation of user $u$ on item $i$ could be modeled as inner products as follows:\par
\begin{equation}
\widehat {{r_{ui}}} = p_u^T{q_i}
\end{equation}
where $\widehat {{r_{ui}}}$ is the estimate of ${r_{ui}}$, ${p_u}$ and ${q_i}$ are column vectors with $f$ values. The layer of $f$-th parameters of all vectors ${p_u}$ and ${q_i}$ is called the $f$-th feature. The feature matrix P and Q can be learned by minimizing the following loss (objective) function:
\begin{equation}
\mathop {\min }\limits_{P,Q} {\sum\limits_{(u,i) \in \kappa } {({r_{ui}} - p_u^T{q_i})} ^2} + \frac{{{\lambda _u}}}{2}\left\| P \right\|_F^2 + \frac{{{\lambda _i}}}{2}\left\| Q \right\|_F^2
\end{equation}
where ${\lambda _u}$,${\lambda _i} > 0$. In order to avoid over-fitting, two regularization terms are added into the loss function. $\left\| . \right\|$ denotes the Frobenius norm.\par

\subsection{UDiff algorithm}
\label{UDiff algorithm}
UDiff algorithm [8] utilizes user-item-tag tripartite graph with resource allocation to recommend new users. The tripartite graph consists of two bipartite graphs: the user-item graph and the user-tag graph, which can be presented by two adjacent matrices $A$ and $A'$. If user $u$ has collected an item $i$, there is an edge between $u$ and $i$, and we set ${\alpha _{{\rm{ui}}}}$ =1, otherwise ${\alpha _{{\rm{ui}}}}$=0. Analogously, if user $u$ has used a tag $t$, we set ${\alpha '_{ut}}$ = 1, otherwise ${\alpha '_{ut}}$ = 0. An example of the process of resource reallocation is illustrated in Figure 1.

\begin{figure}[H]
\centering\includegraphics[width=0.7\textwidth,height=0.25\textheight]{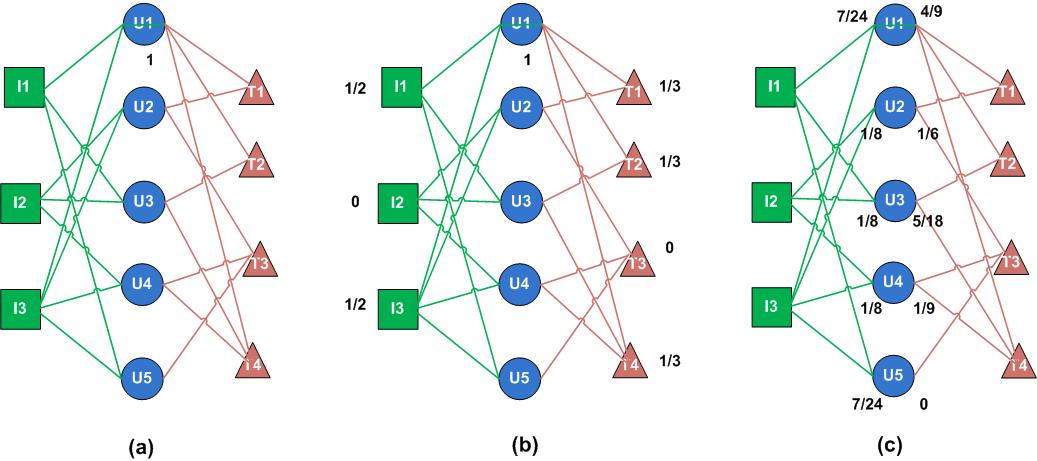}
\caption{The UDiff algorithm at work in the tripartite graph}
\label{fig:1}       
\end{figure}

In the user-item bipartite graph, the resource-allocation process consists of two steps: first from users to items, then back to users. Each user distributes his/her initial resource equally to all the items he/she has collected, and then each item distributes its resource equally to all the users having collected it. Thus, based on user-item bipartite network, the similarity between $u$ and $v$ is:

\begin{equation}
{s_{uv}} = \frac{1}{{k(v)}}\sum\nolimits_{i = 1}^n {\frac{{{\alpha _{ui}}{\alpha _{vi}}}}{{k(i)}}}
\end{equation}

where $k(i)$ is the degree of item $i$ and $k(v)$ is the degree of user $v$ in the user-item bipartite graph. Analogously, considering the diffusion on the user-tag bipartite graph, the tag-based similarity between user $u$ and $v$ is:

\begin{equation}
s{'_{uv}} = \frac{1}{{k'(v)}}\sum\nolimits_{t = 1}^r {\frac{{\alpha {'_{ut}}\alpha {'_{vt}}}}{{k'(t)}}}
\end{equation}

where $k'(t)$ and $k'(v)$ are respectively the degrees of tag $t$ and user $v$ in the user-tag bipartite graph.\par
Finally, the above two diffusion-based similarities are integrated by linear superposition to obtain a final redistribution matrix as:

\begin{equation}
s_{uv}^* = \lambda {s_{uv}} + (1 - \lambda )s{'_{uv}}
\end{equation}

where $\lambda  \in [0,1]$ is a tunable parameter.\par

\section{WUDiff_RMF model}
In this section, we present a novel hybrid collaborative filtering model named WUDiff_RMF, which utilizes the information of similar users from WUDiff algorithm to further improve RMF. And we systematically describe how to model the neighborhood information of users based on tags and ratings as regularization terms to constrain the matrix factorization framework.\par

\subsection{Implicit neighborhood information}
As shown in Fig. 2(a), one user $u1$ has collected item $i1$ with tag $t1$, user $u2$ has collected item $i1$ with tag $t2$ and tag $t3$, and item $i2$ also has been collected by user $u3$ with tag $t3$. If we adopt traditional neighborhood-based CF methods to measure the similarity between users $u2$ and $u3$, they will not be similar at all, for the reason that there is co-rated items. Analogously, due to without co-occurrence tags, user $u1$ and $u2$ also will be dissimilar by the traditional tag-based CF methods. However, user $u1$ and $u2$ have labeled item $i1$ with different tags. In Fig. 2(b), when user Joe and Sam give high ratings to \lq\lq Someone like you \rq\rq and \lq\lq Rolling in the deep \rq\rq respectively, a good recommender system should consider that they have similar preference due to two songs belonged to Adele and the co-occurrence tag. Therefore, compared to traditional similarity measures, UDiff algorithm can be directly applied in extracting the implicit information of similar users based on the degree correlation of the user-item-tag tripartite graph.

\begin{figure}[H]
\centering\includegraphics[width=0.8\textwidth,height=0.3\textheight]{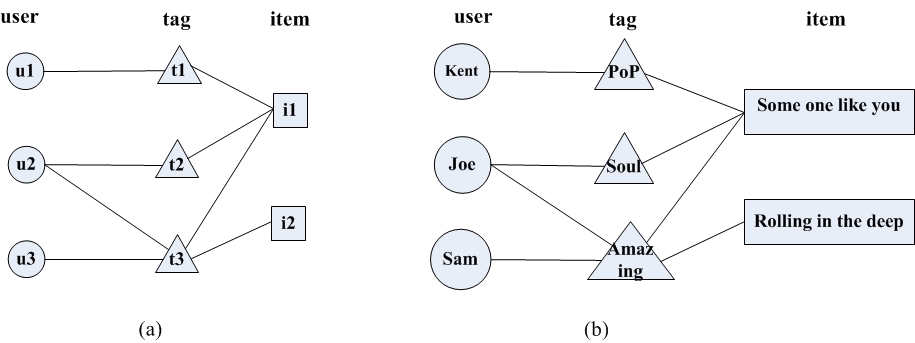}
\caption{The correlation of users in the tripartite graph}
\label{fig:2}       
\end{figure}

However, the UDiff method focuses on the unweighted graphs, and a refined method should take into account the weights of edges in the tripartite network []. Thus, we improve UDiff algorithm by introducing user-item rating as the weights of edges on the user-item bipartite graph and user-tag labeling as the weights on the user-tag relations. Firstly, different users towards various items may differ, so it is necessary to convert individual ratings to a universal scale. Z-score normalization [25] is the most popular rating normalization scheme and considers the spread in the individual rating scales [2]. In user-based methods, the z-score normalization of ${r_{ui}}$ divides the user-mean-centered rating by the standard deviation ${\delta _u}$ the ratings given by item $u$ [2]:

\begin{equation}
h({r_{ui}}) = \frac{{{r_{ui}} - \overline {{r_u}} }}{{{\delta _u}}}
\end{equation}

Furthermore, to compute the weight of user u to tag k conveniently, we adopt an adaptation of Okapi BM25 weighting scheme based on the idea that each user assigns each tag with a score[26,27].

\begin{equation}
w(u,t) = \log \frac{M}{{{n_u}(t)}} \cdot \frac{{tf(u,t) \cdot ({k_1} + 1)}}{{tf(u,t) + {k_1} \cdot (1 - b + b \cdot \frac{{\left| u \right|}}{{avg(U)}})}}
\end{equation}

where M is the number of users, ${n_u}(t)$ is the number of users who has used tag t, $tf(u,t)$ is the number of times user u has annotated items with tag t. $\left| u \right| = \sum\nolimits_{t = 1}^T {{n_u}(t)}$ represents user profile size, $avg(U)$ is the average of all users' profile size, $b$ and $k_1$ are set to the standard values of 0.75 and 2, respectively.\par
According to the above analysis, the resource allocation process of the WUDiff is formulated as£º

\begin{equation}
{ws_{uv}} = \frac{1}{{k(v)}}\sum\nolimits_{i = 1}^n {\frac{{{\alpha _{ui}}{\alpha _{vi}}h({r_{ui}})h({r_{vi}})}}{{k(i)}}}
\end{equation}
\begin{equation}
ws{'_{uv}} = \frac{1}{{k'(v)}}\sum\nolimits_{t = 1}^r {\frac{{\alpha {'_{ut}}\alpha {'_{vt}}w(u,t)w(v,t)}}{{k'(t)}}}
\end{equation}
\begin{equation}
ws_{uv}^* = \lambda  \cdot w{s_{uv}} + (1 - \lambda ) \cdot w{s'_{uv}}
\end{equation}

\subsection{Regularization and optimization}
In this section, we will incorporate additional sources of information from the user-item-tag tripartite network to improve the prediction accuracy of the RMF approach. WUDiff can obtain the neighborhood information by utilizing the information of ratings and tags, and thus can capture the implicit information of similar users based on the degree correlation of the tripartite network, not just ratings or tags. Since the behavior of users always resembles their neighbors in taste, we integrate similar user regularization term to the loss function in our model, which measures the difference between the latent feature vector of a user and those of his/her similar users from WUDiff. The mechanism of WUDiff_RMF is shown in Fig.3.

\begin{figure}[H]
\centering\includegraphics[width=0.8\textwidth,height=0.3\textheight]{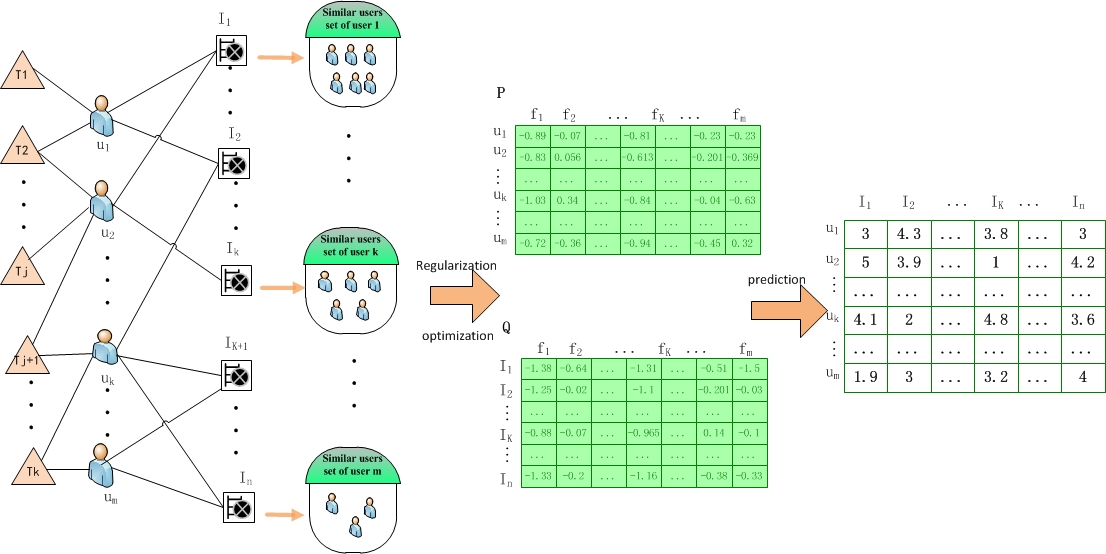}
\caption{The mechanism of WUDiff_RMF}
\label{fig:3}       
\end{figure}

There are a set I of items and a set U of users. An user group denoted by S(u) is a \lq\lq k-nearest-neighbors\rq\rq  set of user u obtained by WUDiff_RMF approach, and S(u) is represented as following:

\begin{equation}
S(u) = \{ u_1^{{w_{u,1}}},u_2^{{w_{u,2}}},...,u_k^{{w_{u,k}}}\}
\end{equation}

where $k$ is the number of users in S(u); ${u_k}$ is a user, and ${w_{u,k}}$ is the similarity user $u$ and user $u_k$.\par
A user-factors vector ${p_u}$ of the user $u$ is a vector of real numbers in the closed interval [0,1], defined by

\begin{equation}
{p_u} = ({p_{u,1}},...,{p_{u,t}},...,{p_{u,f}})
\end{equation}

where $f$ is the number of feature vectors, and ${p_{u,t}}$ is the typicality value of user $u$ in the feature vector $t$.
If user $u$ is similar to user $v$, the ideal property is to minimize the distance between latent vectors ${p_u}$ and ${p_v}$. Hence,to improve recommender systems, we utilize regularization term $\left\| {{p_u} - {p_v}} \right\|_F^2$ employed in Equation (2) to utilize implicit user information from WUDiff.\par
The range of datasets is different, for example, the movie rating is range from 0 to 5. Thus, to learn parameters in a more convenient way, we change the raw rating ${R_{ui}}$ by adopting the function $f(x) = x/{R_{\max }}$, where $R_{\max}$ is the maximum of ratings in the RSs. In Movielens, due to ${R_{\max }} = 5$, an interval [0,5] will be changed into [0,1]. In addition, the logistic function $g(x) = 1/(1 + \exp ( - x))$ is employed to map the inner product of ${p_u}$ and ${q_i}$ into the interval [0,1]. Accordingly, the objective function of this approach can be formulated as:

\begin{equation}
\begin{array}{l}
L = \mathop {\min }\limits_{P,Q} {\sum\limits_{(u,i) \in \kappa } {({r_{ui}} - g(p_u^T{q_i}))} ^2} + \frac{\alpha }{2}\sum\limits_{u = 1}^m {\sum\limits_{v \in S(u)} {ws_{uv}^* \cdot \left\| {{p_u} - {p_v}} \right\|_F^2} } \\
\\
+\frac{{{\lambda _u}}}{2}\left\| P \right\|_F^2 + \frac{{{\lambda _i}}}{2}\left\| Q \right\|_F^2
\end{array}
\end{equation}

where $\alpha > 0$, $\alpha$ is the regularization parameter, which balances the contribution of neighborhood information of users from WUDiff to the recommendation performance of the model. ${ws_{uv}^*}$ indicates the similarity between user $u$ and user $v$. $S(u)$ is the set of users who are similar to user $u$ by WUDiff.\par
For each observed rating ${r_{ui}}$, we can minimize the above objective function by performing Stochastic Gradient Descent (SGD) to learn the latent parameters:
\begin{equation}
\begin{array}{l}
{p_u} \leftarrow {p_u} + {\gamma _1}(g'(p_u^T{q_i}){e_{ui}}{q_i} - \alpha \sum\limits_{v \in S(u)} {ws_{uv}^* \cdot\left( {{p_u} - {p_v}} \right)} \\
 + \alpha \sum\limits_{v \in S(u)} {ws_{uv}^* \cdot\left( {{p_v} - {p_u}} \right)}  - {\lambda _u}{p_u}),\\
 \\
 {q_i} \leftarrow {q_i} + {\gamma _2}(g'(p_u^T{q_i}){e_{ui}}{p_u} - {\lambda _i}{q_i})
\end{array}
\end{equation}
where ${e_{ui}} = {r_{ui}} - g(p_u^T{q_i})$, and the derivative of logistic function g(x) is $g'(x) = \exp ( - x)/{(1 + \exp ( - x))^2}$. ${\gamma _1}$ and ${\gamma _2}$ are the learning rates.\par

Different from previous approaches that incorporate localized relationships into RMF model, WUDiff_RMF combines neighborhood information of users on the degree correlation of the weighted tripartite graph based on ratings and tags, instead of using co-occurrence properties of tags or ratings for finding the set of neighbors. The learning algorithm of our model is described in algorithm 1:
\begin{algorithm}[H]
  \caption{WUDiff_RMF}
  \begin{algorithmic}[1]
   \renewcommand{\algorithmicrequire}{ \textbf{Input:}} 
    \Require
     train set ${D_T}$, Test set ${D_V}$, factor dimensionality $f$, user_num, R(u), S(u)
     \renewcommand{\algorithmicensure}{ \textbf{Output:}} 
     \Ensure
      Learned WUDiff_RMF model parameters
      \State Randomly initialize feature parameters;
      \Repeat
       \For {${\rm{ }}u = 1;u <  = user\_num;{\rm{ }}u +  + {\rm{ }}$}
       \For {${\rm{ }}all{\rm{ }}\quad  i \in R\left( u \right)$}
         \State ${p_u} \leftarrow {p_u} + {\gamma _1}(g'(p_u^T{q_i}){e_{ui}}{q_i} - \alpha \sum\limits_{v \in S(u)} {ws_{uv}^* \cdot\left( {{p_u} - {p_v}} \right)}$
         \State $ + \alpha \sum\limits_{v \in S(u)} {ws_{uv}^* \cdot\left( {{p_v} - {p_u}} \right)}  - {\lambda _u}{p_u})$
         \State ${q_i} \leftarrow {q_i} + {\gamma _2}(g'(p_u^T{q_i}){e_{ui}}{p_u} - {\lambda _i}{q_i})$
      \EndFor
      \EndFor
    \State update;
    \State calculate RMSE on ${D_V}$;
    \Until RMSE on ${D_V}$ does not improve.
    \label{code:recentEnd}
  \end{algorithmic}
\end{algorithm}

\section{Experiments}
\label{sec:5}
In this section, we report the results from experiments conducted to compare the WUDiff_RMF with seven existing recommendation methods by using well-known benchmark datasets. Our experiments aim to answer the following questions:
\begin{enumerate}
\item  How does the parameter $\lambda$ of WUDiff algorithm affect the WUDiff_RMF method?
\item  How does the number of nearest neighbor $k$ affect the recommendation quality?
\item  How does the parameter $\alpha$ affect the recommendation results?
\item  Is CogTime RMF more efficient than seven existing recommendation methods?
\item  Can our method obtain good performance even if users have very few ratings and few tags?
\end{enumerate}

\subsection{Description of Datasets}
We adopt four real-world datasets, Delicious, Last.fm, DBLP, and Movielens, to evaluate the effectiveness of our recommendation algorithm. The first two datasets are released in the conference of HetRec 2011[28]. Delicious consists of 1867 users, 69,926 URLs and 53,388 tags. Last.fm consists of 1892 users, 17,632 artists and 11,946 tags. The third data set DBLP contains 6815 authors' ratings on 78475 papers with 81858 unique venues and authors of all papers from an academic network[23,29]. The three datasets have binary ratings. The fourth dataset Movielens, which has 1857 users' ratings on 4721 items(scale from 0.5 to 5) with 8288 unique tags[23,28], is also released in the framework of HetRec 2011.

\subsection{Evaluation metrics}
We employ two widely used metrics, Mean Absolute Error (MAE) and Root Mean Square Error (RMSE), to investigate the prediction quality of our proposed WUDiff_RMF model in comparison with other collaborative filtering methods.\par
MAE is defined as:

\begin{equation}
MAE = \sum\limits_{(u,i,{r_{ui}}) \in {T_E}} {\left| {{r_{ui}} - \widehat {{r_{ui}}}} \right|/\left| {{T_E}} \right|}
\end{equation}

where ${T_E}$ denotes the number of tested ratings.\par
RMSE is computed as follows:

\begin{equation}
RSME = \sqrt {\sum\limits_{(u,i,{r_{ui}}) \in {T_E}} {{{({r_{ui}} - \widehat {{r_{ui}}})}^2}/\left| {{T_E}} \right|} }
\end{equation}

Observably lower MAE and RMSE correspond to higher prediction accuracy.

\subsection{Comparisons}
In the experiments, to show the performance of our proposed WUDiff_RMF, we adopt two matrix factorization recommendation methods as baselines for the comparison, which include RMF and PMF without tagging data. Moreover, we compare our algorithm with five state-of-the-art algorithms with tagging: CTR, FM, RLFM, SIM and FCR-r.
\begin{itemize}
\item RMF: It has been described in Section 3.1 and uses only the user-rating matrix to generate recommendations.
\item PMF [14]: It models user-item rating matrix by matrix factorization, including adaptive priors over the item and user feature vectors that can be used to control model complexity automatically.
\item Collaborative topic regression (CTR) [30]: It combines matrix factorization and probabilistic topic modeling for the recommendation.
\item Factorization machine (FM) [31]: It combines the flexibility of feature engineering with the superiority of factorization models.
\item Regression latent factor model (RLFM) [32]: It is a regression based latent factor model by incorporating past interactions and the features of users and items.
\item Similarity based method (SIM) [33]: It incorporates tag preferences in the context of an item by adopting the SVM regression for recommendation.
\item Feature-Centric Recommendation(FCR-r) [23]: It transforms the item ratings into feature ratings and learns the global weighting for features by a regression model.
\end{itemize}
The results of PMF, CTR, FM, RLFM, SIM and FCR-r are cited from [23]. To a fair comparison, we follow the experimental settings as [23], and the results of RMF and WUDiff_RMF are obtained by our careful implementations.\par

\subsection{Experimental Results and analysis}
We conduct a series of experiments on four datasets to demonstrate the advantage of our method relative to others. To assess the performance of WUDiff_RMF, we apply the standard 10-fold cross-validation for all datasets and methods. The random selection is carried out 10 times independently, then mean of MAE and RMSE are calculated as the final results. For RMF, we use the setting ${\lambda _u} = {\lambda _v} = 0.01$.

\subsubsection{Impact of the parameter $\lambda $}
\label{sec:1}
In WUDiff, as a tunable parameter, $\lambda$ determines how much the information of tag should be incorporated into diffusion-based similarity in the user-item-tag tripartite network. $\lambda = 1$ and $\lambda = 0$ correspond to the cases for pure user-item and user-tag diffusions. To test the effect of the parameter ¦Ë on the recommendation accuracy of WUDiff_RMF, we conduct experiments by setting $\lambda$ from 0 to 1 on different datasets.\par
\begin{figure}[H]
\subfigure[$\lambda $ vs. RMSE, Delicious] {\includegraphics[width=0.5\textwidth]{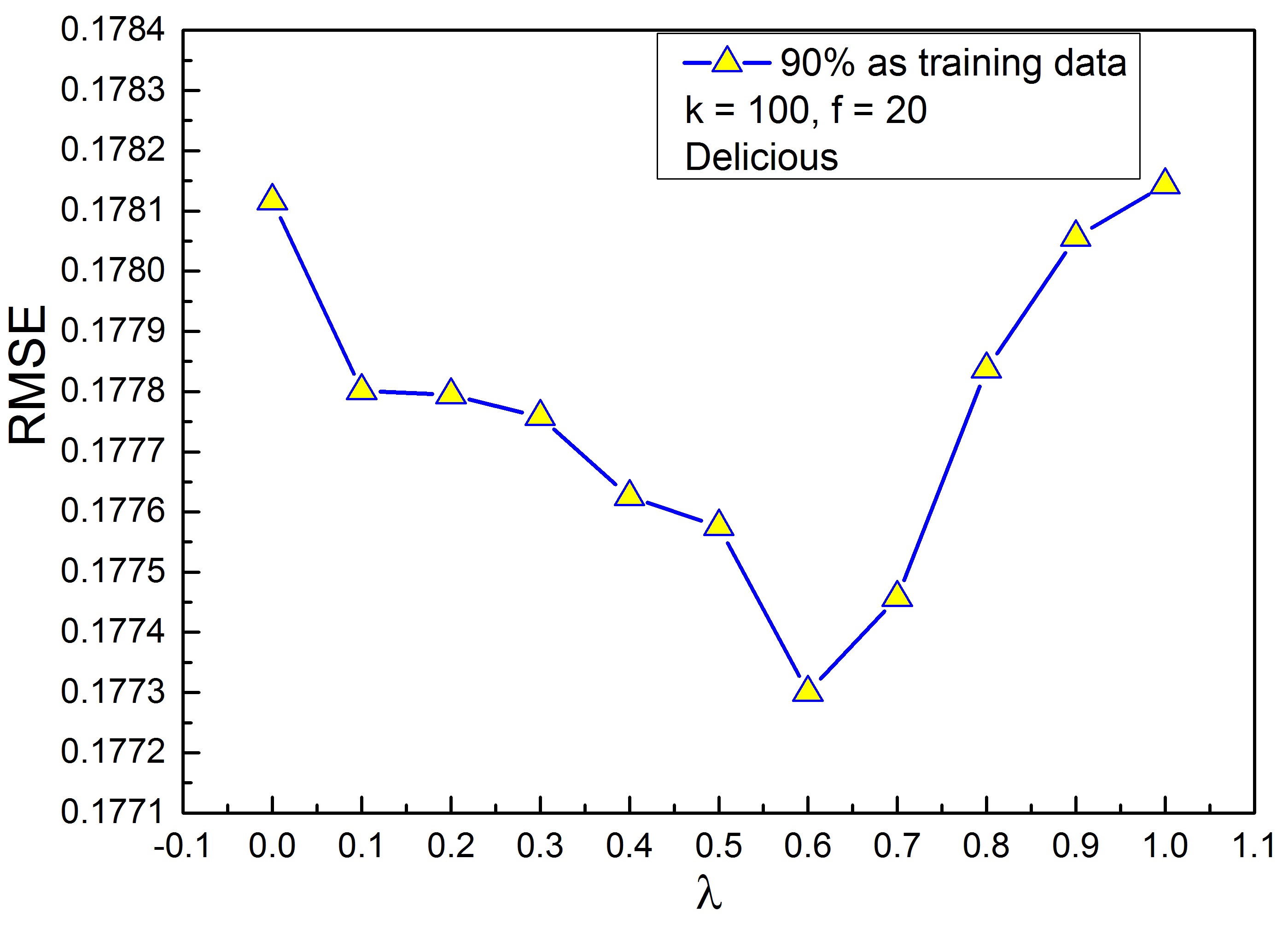}}
\subfigure[$\lambda $ vs. RMSE, Last.fm] {\includegraphics[width=0.5\textwidth]{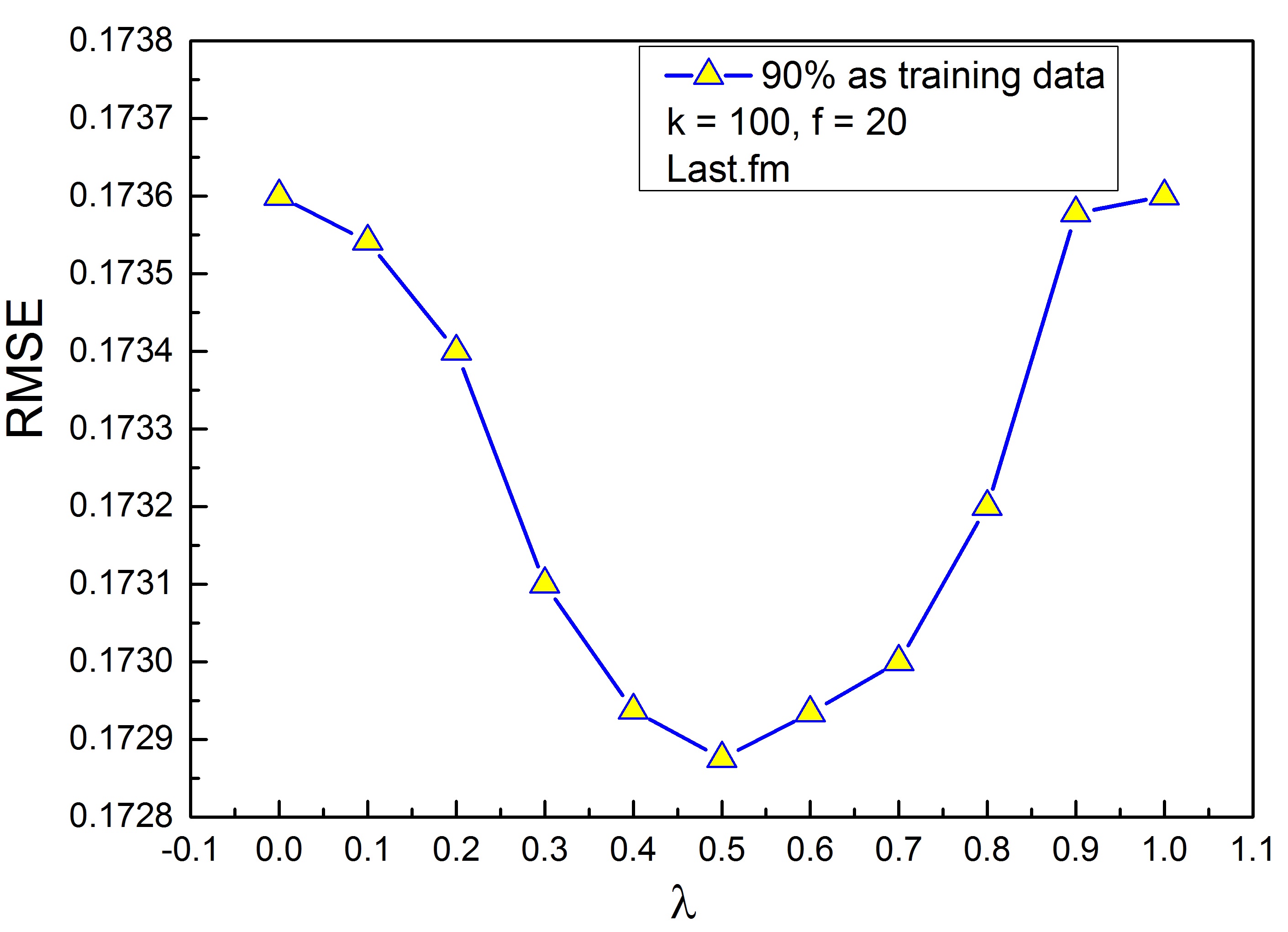}}
\subfigure[$\lambda $ vs. RMSE, DBLP] {\includegraphics[width=0.5\textwidth]{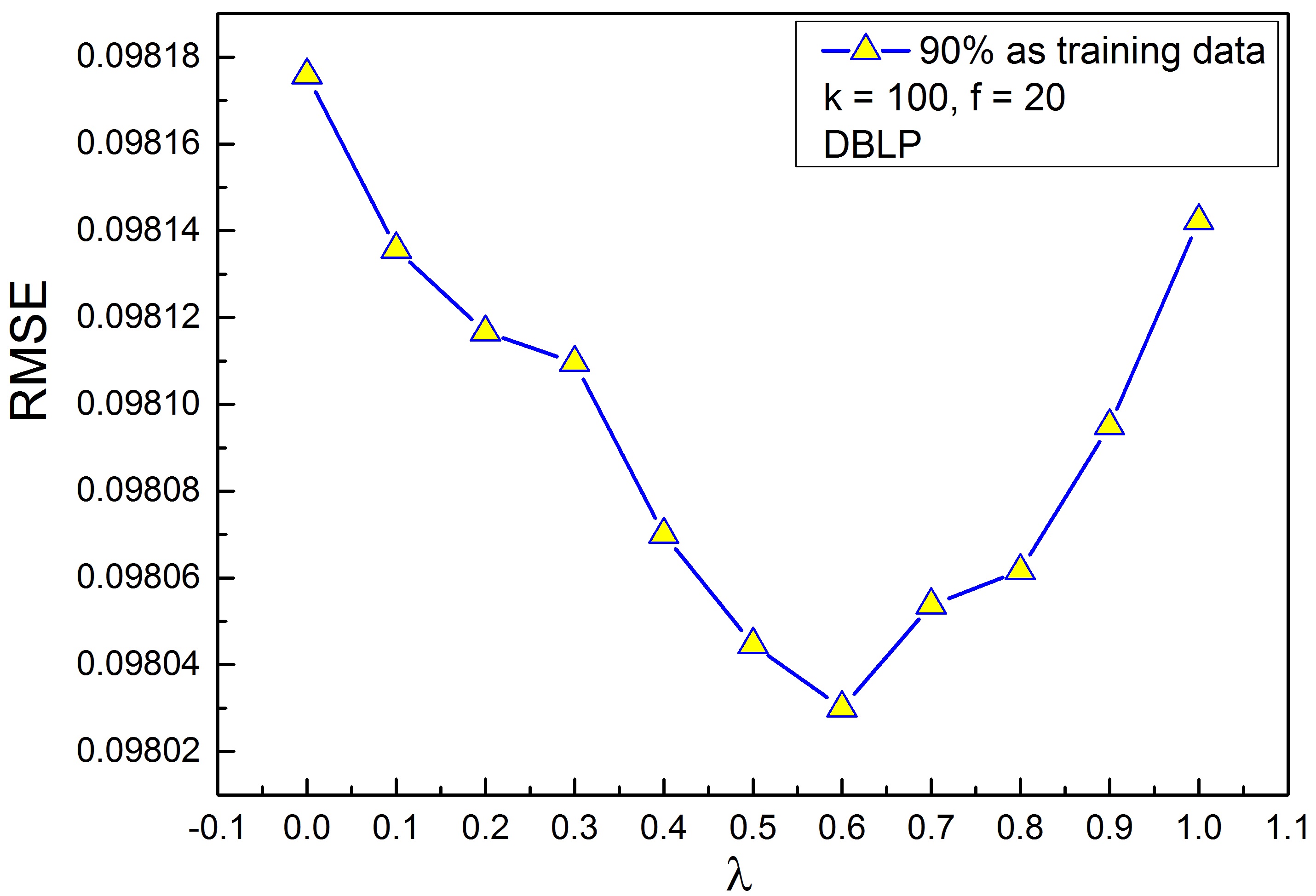}}
\subfigure[$\lambda $ vs. RMSE, Movielens] {\includegraphics[width=0.5\textwidth]{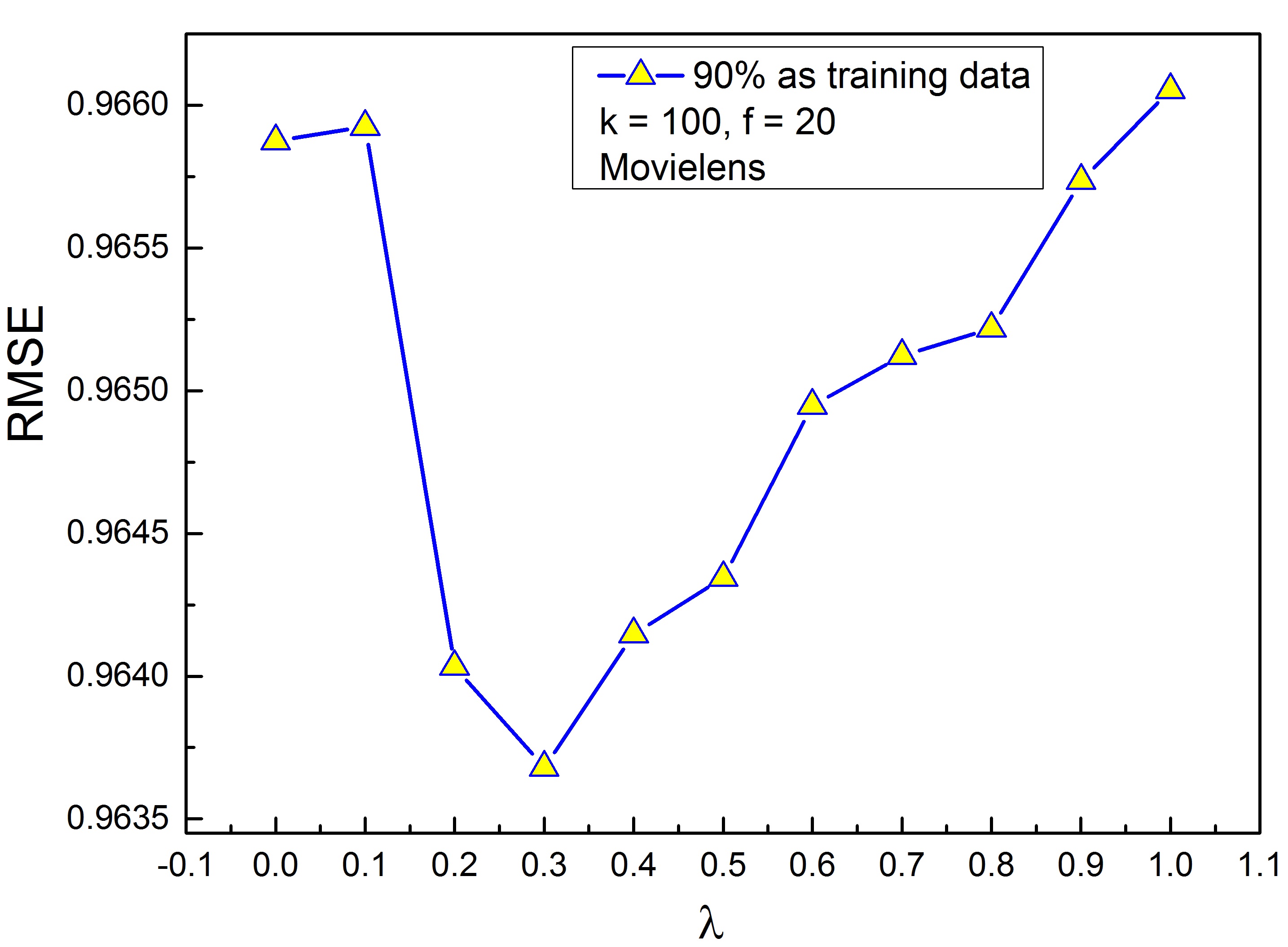}}
\caption{Impact of Parameter $\lambda$ }
\label{fig:4}       
\end{figure}
As shown in Fig.4, to compare the $\lambda = 0$ or $\lambda = 1$ case, we find that the balanced relation between taggings and ratings can help improving the algorithmic accuracy. As $\lambda$ increases, the RMSE value decreases at first, but when $\lambda$ goes up a certain threshold like 0.6 on Delicious dataset, the RMSE value increases steeply with further increase of the value of $\lambda$. The best parameter $\lambda$ setting for other three datasets are 0.5, 0.6, and 0.3 respectively.\par

\subsubsection{Impact of the parameter $k$}
We analyze the performance of our model when varying the number of nearest neighbor $k$ on four datasets. Different $k$ will lead different recommendation accuracies. Fig.5 shows following observations. With the increment of $k$, the prediction accuracy improves at first, but when $k$ goes up a certain threshold like 40 on Delicious dataset, the prediction accuracy decreases. Accordingly, the thresholds of parameter $k$ are 40 on Last.fm, 30 on DBLP and 40 on Movielens.
\begin{figure}[H]
\subfigure[$k$ vs. RMSE, Delicious] {\includegraphics[width=0.5\textwidth]{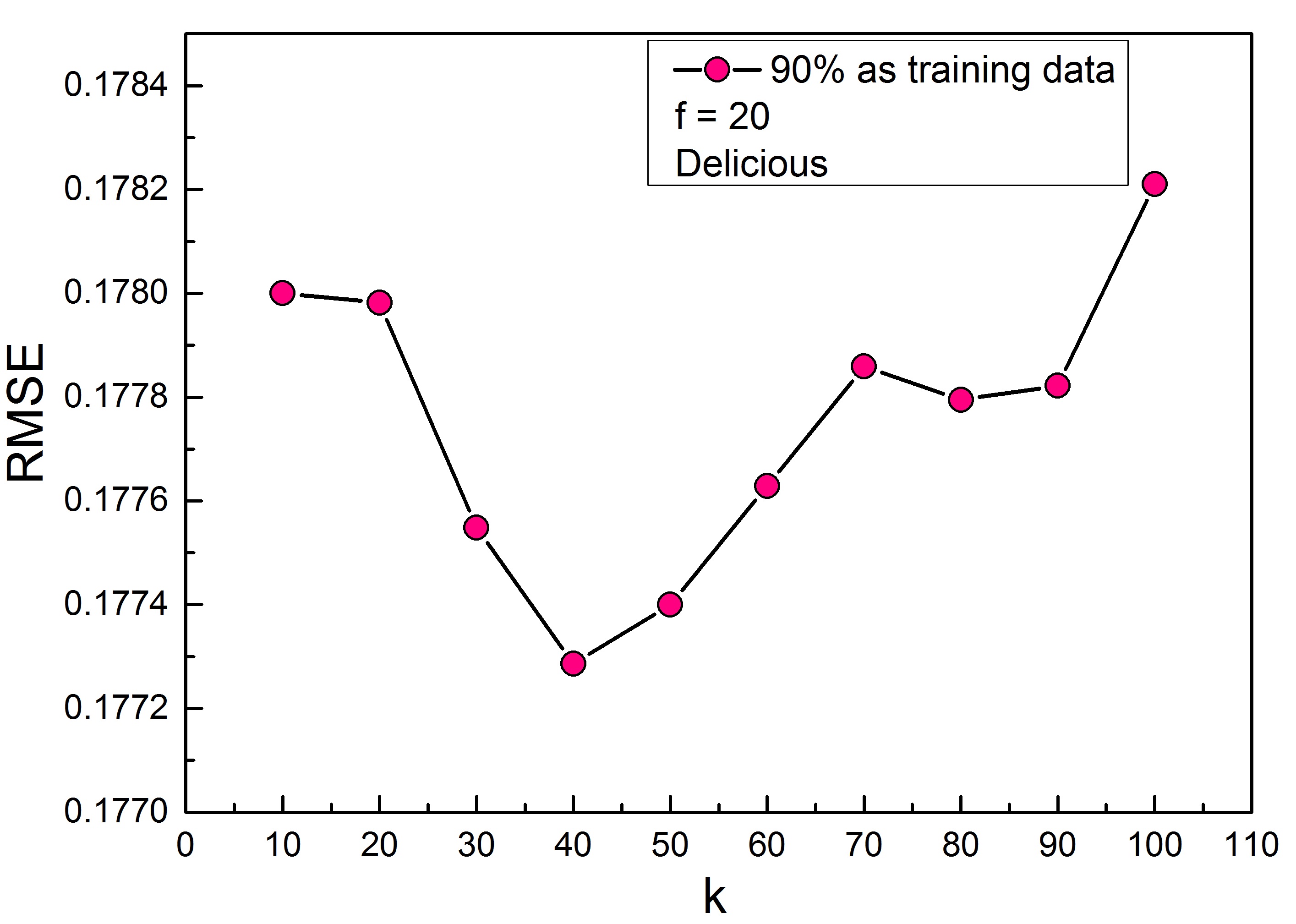}}
\subfigure[$k$ vs. RMSE, Last.fm] {\includegraphics[width=0.5\textwidth]{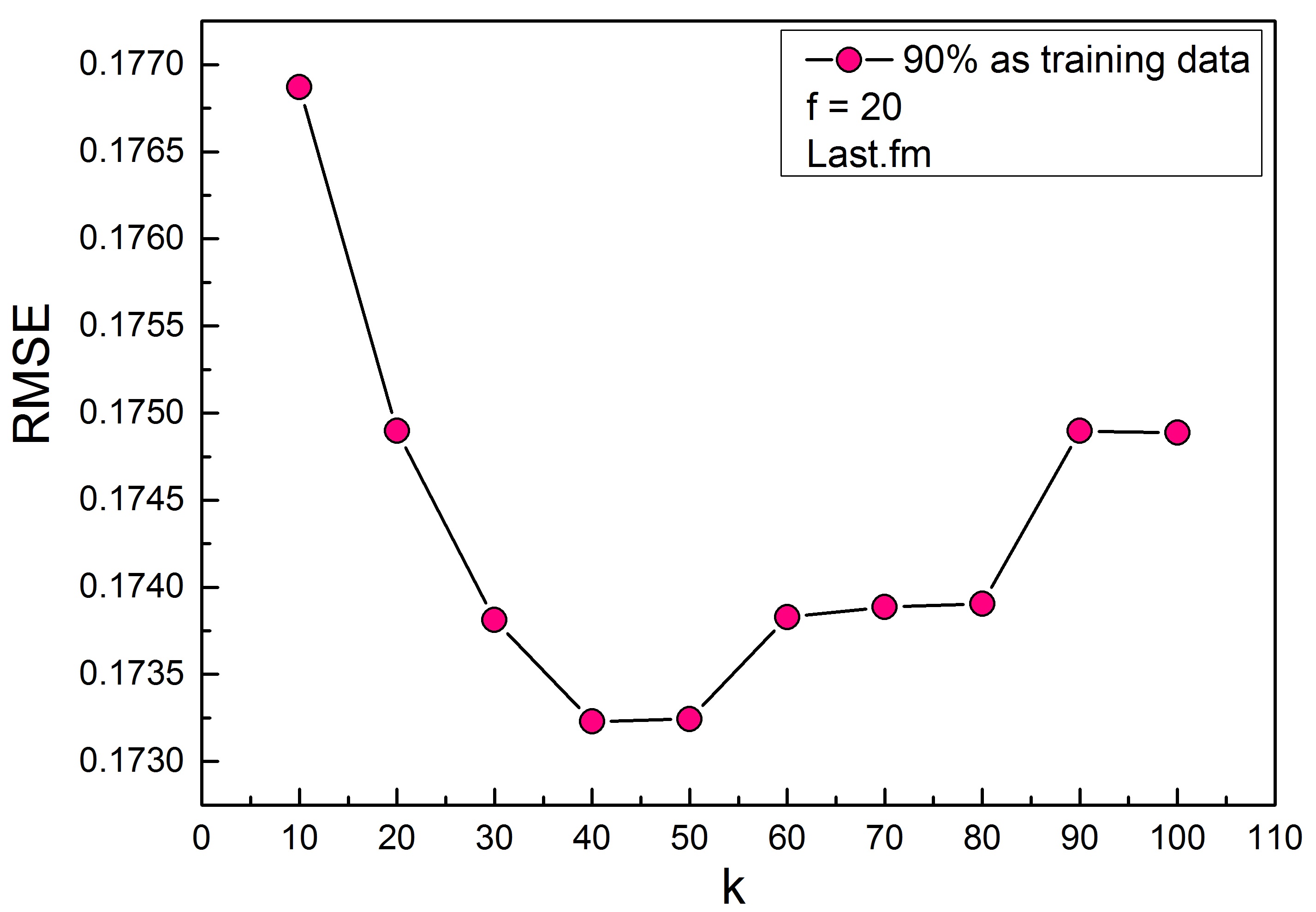}}
\subfigure[$k$ vs. RMSE, DBLP] {\includegraphics[width=0.5\textwidth]{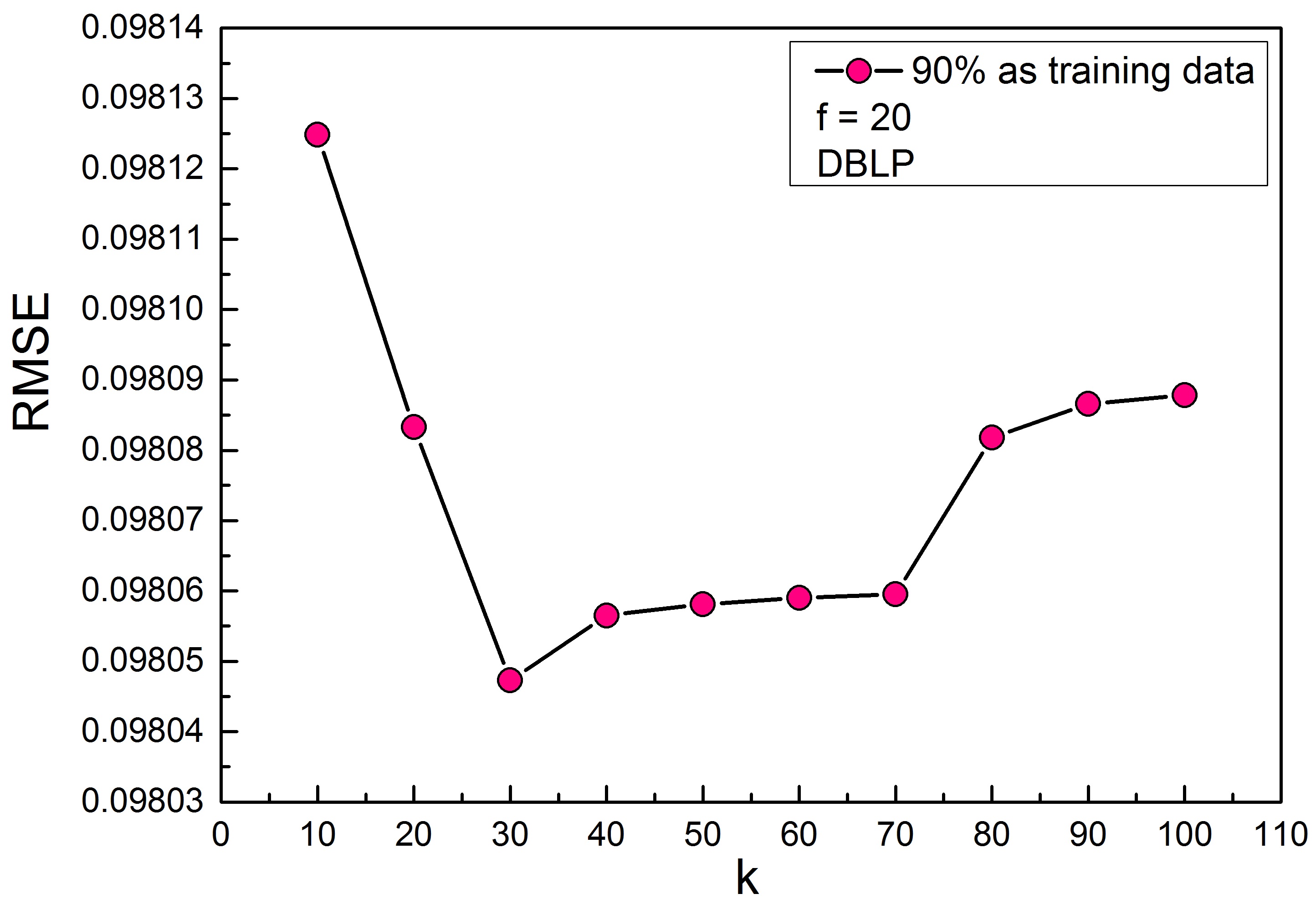}}
\subfigure[$k$ vs. RMSE, Movielens] {\includegraphics[width=0.5\textwidth]{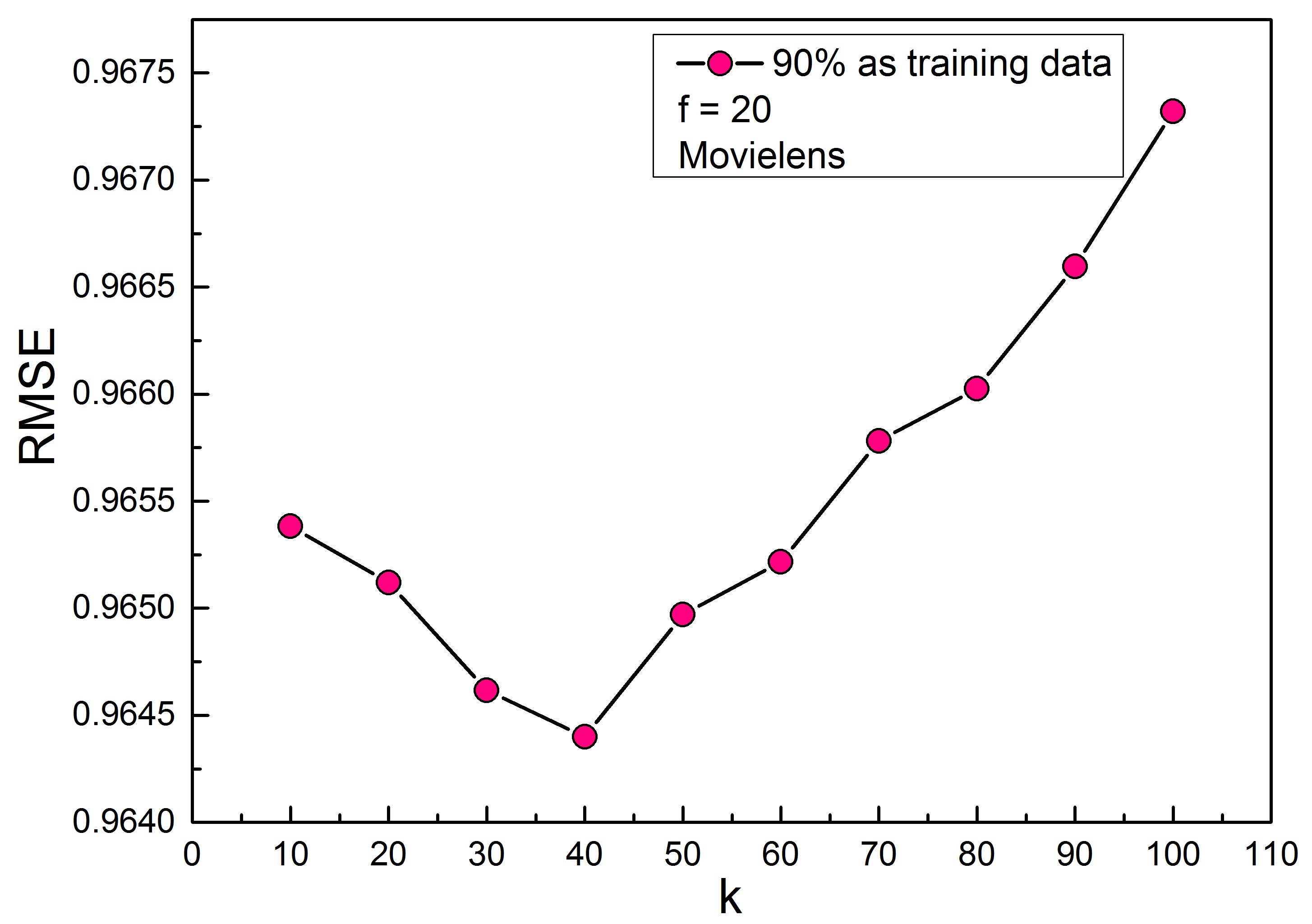}}
\caption{Impact of Parameter $k$ }
\label{fig:5}       
\end{figure}
Although $k$ increases, recommendation accuracy would be improved in KNN-based CF methods [2,6]. However, in WUDiff_RMF, when $k$ is too small, the set of \lq\lq neighbors\rq\rq  is not enough to improve recommendation accuracy. As $k$ increase, there are not enough qualified \lq\lq neighbors\rq\rq  for each user that will affect the performance of RMF in capturing overall structure.

\subsubsection{Impact of the parameter $\alpha$}
In WUDiff_RMF, the parameter $\alpha$ controls the proportion of effect between rating preference and neighborhood information from the user-item-tag tripartite network when training model. In order to check the effect of the parameter $\alpha$ for recommending, we compare the RMSE of our model for the different ranges of parameter $\alpha$ ¦Á on all datasets.\par
\begin{figure}[H]
\subfigure[$\alpha$ vs. RMSE, Delicious] {\includegraphics[width=0.5\textwidth]{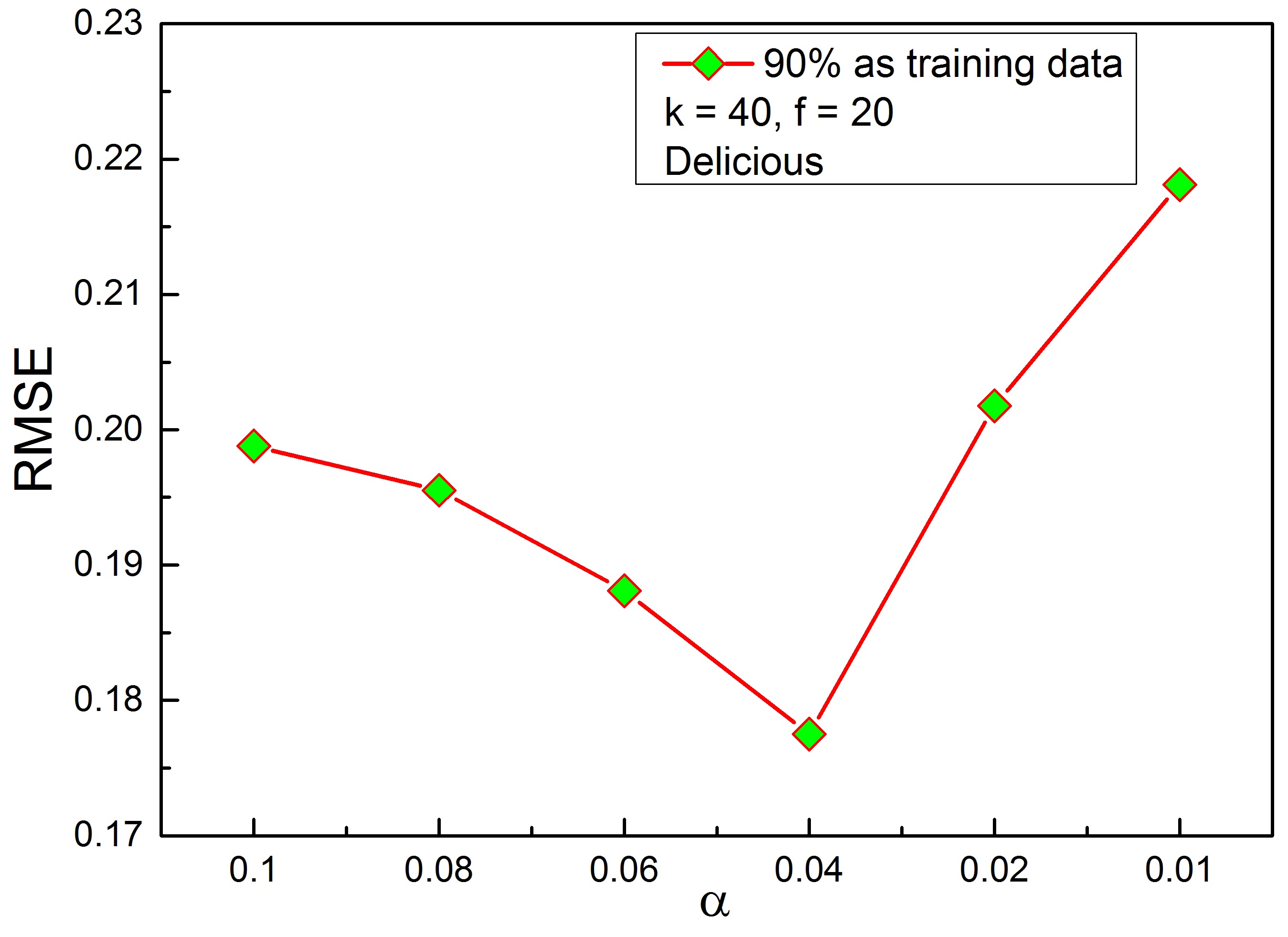}}
\subfigure[$\alpha$ vs. RMSE, Last.fm] {\includegraphics[width=0.5\textwidth]{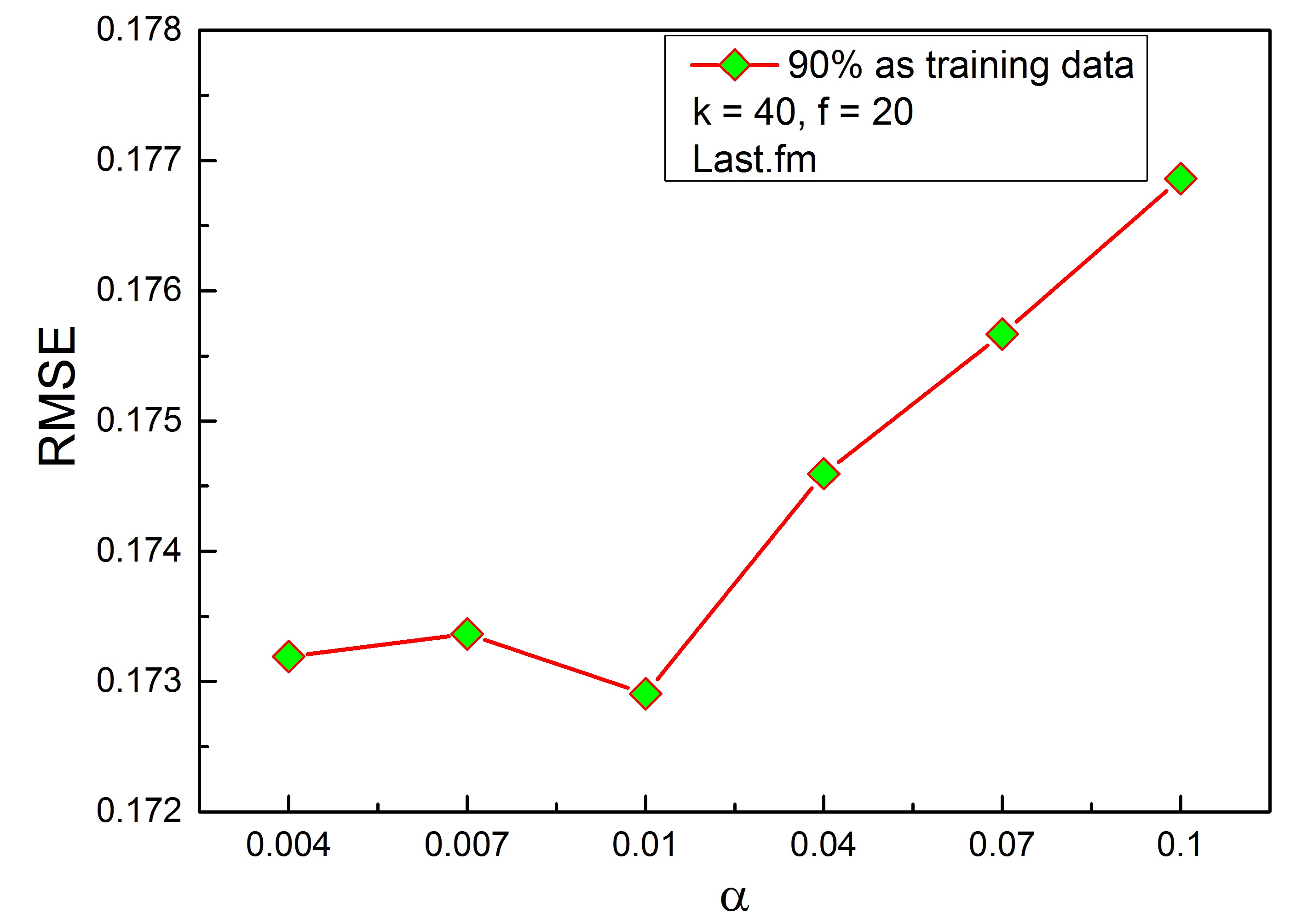}}
\subfigure[$\alpha$ vs. RMSE, DBLP] {\includegraphics[width=0.5\textwidth]{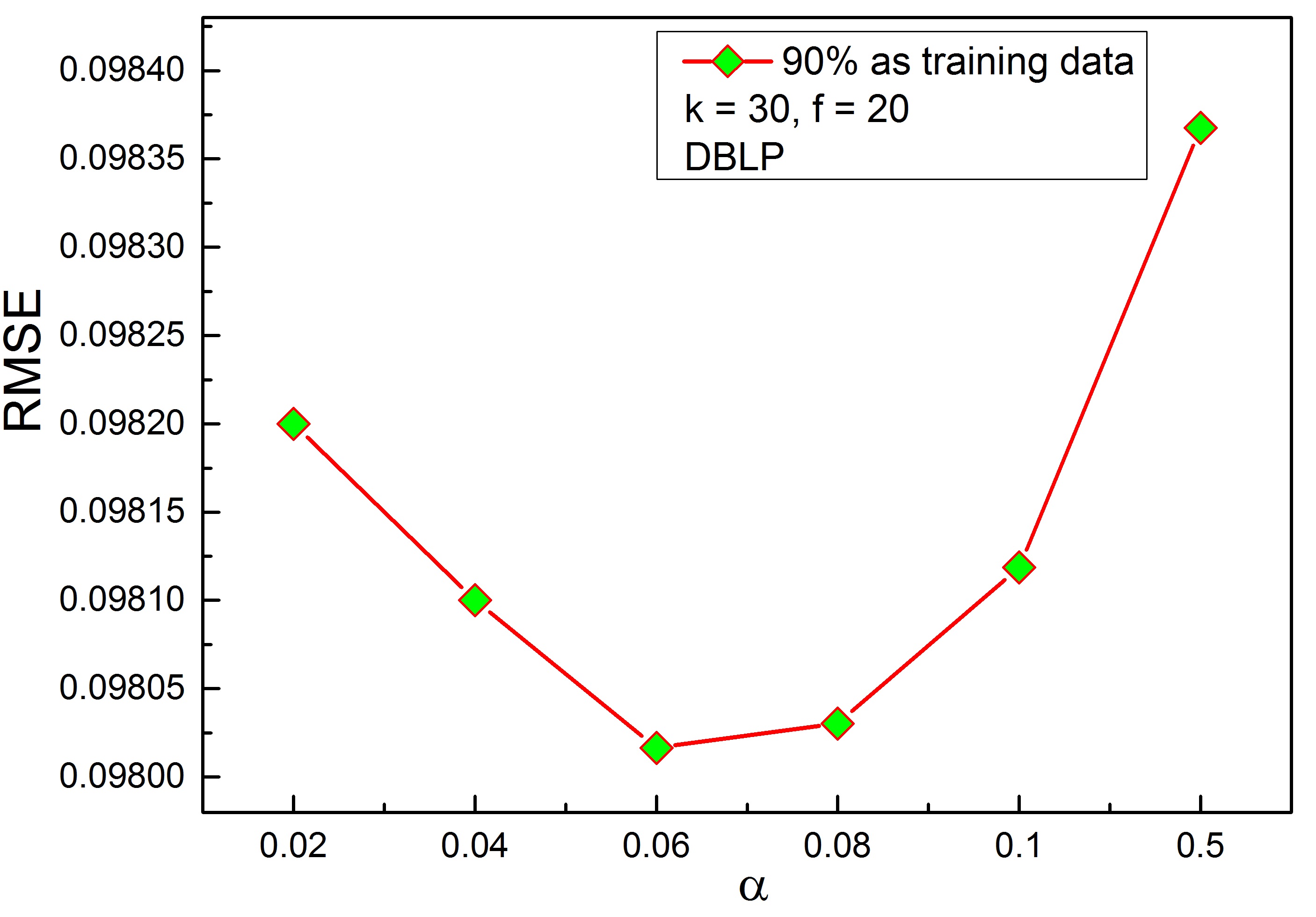}}
\subfigure[$\alpha$ vs. RMSE, Movielens] {\includegraphics[width=0.5\textwidth]{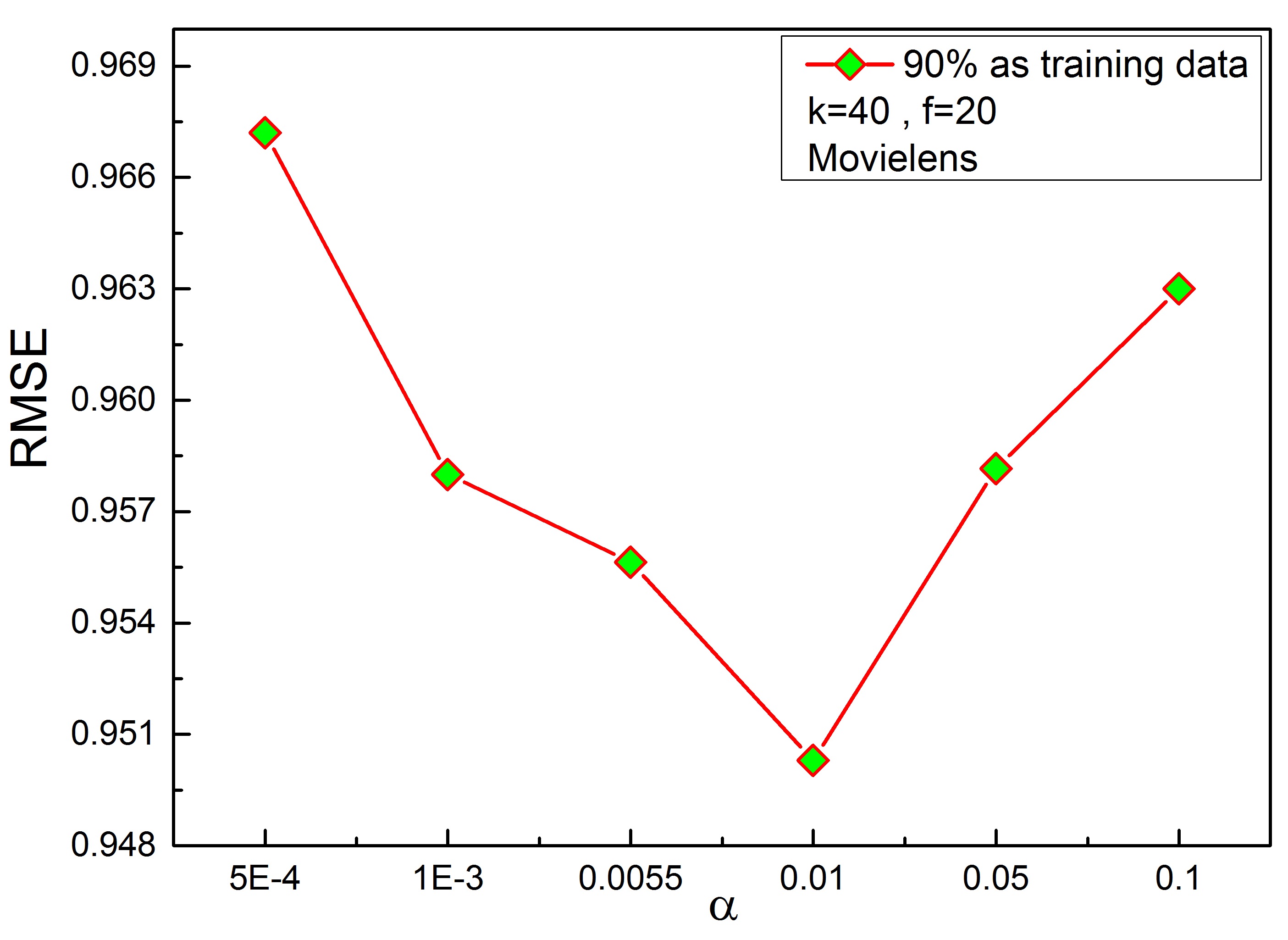}}
\caption{Impact of Parameter $\alpha$ }
\label{fig:5}       
\end{figure}
From the results presented in Fig.6, we can observe that incorporating neighborhood information of users can actually enhance the performance of recommendation. If we set $\alpha$ as 0, it means we just employ the user-item rating matrix for making the recommendation. As $\alpha$ increases, the RMSE value decreases at first, but when $\alpha$ goes below a certain threshold like 0.04 on Delicious dataset, the RMSE value goes up with further increase of the value of $\alpha$. The best value of $\alpha$ on Last.fm, DBLP, Movielens is 0.01,0.06, and 0.01.\par

\begin{table*}[htbp]
\scriptsize
\caption{MAE and RMSE comparison on four datasets}
\label{tab:2}       
\begin{tabular}{llllllllll}
\hline\noalign{\smallskip}
Dataset	 & Metrics & RMF & PMF & CTR & FM & RLFM & SIIM & FCR-r &WUDiff\\
	 &  &  &[23]  &[23]  &[23]  & [23] & [23] & [23] &_RMF\\
\noalign{\smallskip}\hline\noalign{\smallskip}
Delicious & MAE &0.1891 & 0.8081 & 0.7431 &	0.2906  & 0.3978 & 0.3872 & 0.1513 & 0.1482\\
  &  &$\pm$0.0225 & $\pm$0.0048 & $\pm$0.0005 & $\pm$0.0020 & $\pm$0.0010 & $\pm$0.0011 & $\pm$0.0009 & $\pm$0.0008\\
  & RMSE &0.3732 & 0.8907 &	0.7844 & 0.3551 & 0.4182 & 0.4001 & 0.2176 & 0.1775\\
  &  &$\pm$0.0378 & $\pm$0.0046 & $\pm$0.0004 & $\pm$0.0017 & $\pm$0.0010 & $\pm$0.0008 & $\pm$0.0011 & $\pm$0.0014\\
Last.fm  & MAE &0.185 & 0.3179 & 0.4084 & 0.2534 & 0.2235 & 0.2941 & 0.1066 & 0.0762\\
  &  &$\pm$0.0024 & $\pm$0.0022 & $\pm$0.0025 & $\pm$0.0030 & $\pm$0.0015 & $\pm$0.0013 & $\pm$0.0016 & $\pm$0.0022\\
  & RMSE &0.3205 & 0.4449& 0.5078& 0.3239 & 0.3208&	0.3269&	0.1868 & 0.1729\\
  &  &$\pm$0.0198 & $\pm$0.0040& $\pm$0.0025 & $\pm$0.0022 & $\pm$0.0014& $\pm$0.0013& $\pm$0.0020 & $\pm$0.0032\\
DBLP  & MAE &0.1827 & 0.3800 & 0.3653 &	0.1167 & 0.1930 & 0.3032 & 0.0841 & 0.0367\\
  &  &$\pm$0.0040 & $\pm$0.0020 & $\pm$0.0018 &	$\pm$0.0020 & $\pm$0.0007 & $\pm$0.0003& $\pm$0.0003 & $\pm$0.0004\\
  & RMSE &0.2871 & 0.5060 & 0.4943&	0.1821& 0.2297 & 0.3064& 0.1064& 0.0980\\
  &  &$\pm$0.0018 & $\pm$0.0021 & $\pm$0.0021 &	$\pm$0.0023& $\pm$0.0007& $\pm$0.0003& $\pm$0.0004& $\pm$0.0006\\
Movielens & MAE &1.0748 & 0.8622 & 0.8315&	0.9467& 0.8056 & 0.7616& 0.7208& 0.7280\\
  &  &$\pm$0.0187 & $\pm$0.0197& $\pm$0.0156& $\pm$0.0194& $\pm$0.0145 & $\pm$0.0123& $\pm$0.0125& $\pm$0.0045\\
  & RMSE &1.4495 & 1.1271& 1.0880&	1.2049&	1.0662&	1.0137& 0.9724& 0.9502\\
  &  &$\pm$0.0182 & $\pm$0.0247& $\pm$0.0191 & $\pm$0.0229&	$\pm$0.0215& $\pm$0.0190& $\pm$0.0201& $\pm$0.0027\\
\noalign{\smallskip}\hline
\end{tabular}
\end{table*}

\subsubsection{Prediction performance of WUDiff_RMF}
As mentioned in Section 5.3, we adopt two baseline methods and five state-of-the-art models to compare with WUDiff_RMF on four datasets in terms of RMSE and MAE, in which we set $f$=20. Note that the results of PMF, CTR, FM, RLFM, SIM and FCR-r are directly cited from [13]. From Table 2, we can notice that our method outperforms other approaches on all datasets. For example, on average, WUDiff_RMF improves FCR-r by 18.43\%, 7.45\%, 7.89\%, and 2.28\% respectively, in terms of RMSE, on the four datasets. Although FCR-r performs better than our method in Movielens in terms of MAE, WUDiff_RMF is more powerful in terms of RMSE. Since RMSE is more indicative than MAE [34], our method still owns best performance.\par
\begin{table*}[htbp]
\footnotesize
\caption{Statistical significance of prediction accuracy improvements}
\label{tab:3}       
\begin{tabular}{llllll}
\hline\noalign{\smallskip}
Method	 & t-Test & Delicious & Last.fm & DBLP & Movielens \\
\noalign{\smallskip}\hline\noalign{\smallskip}
RMF with user-item & MAE &t=22.24 & t=41.78 & t=31.78 & t=12.63	\\
bipartite network  &  &p=1.71E-6 & p=7.41E-8 & p=2.92E-6 & p=p=2.76E-5 \\
 & RMSE &t=5.29 & t=4.23 & t=18.78 & t=5.37 \\
 &  &p=0.0016 & p=0.0041 & p=2.37E-5 & p=0.0015\\
RMF with user-tag & MAE &t=23.94 & t=21.83 & t=19.71 & t=4.89	\\
bipartite network  &  &p=1.18E-6 & p=1.87E-6 &p=1.95E-5 & p=0.0022\\
 & RMSE &t=4.28 & t=7.02 & t=23.53 & t=5.42 \\
 &  &p=0.0039 & p=4.52E-4 & p=9.67E-6 & p=0.0014\\
\noalign{\smallskip}\hline
\end{tabular}
\begin{tablenotes}
\item *Significance at 95\%.
\end{tablenotes}
\end{table*}

In order to further validate the statistical significance of our experiments, we perform the paired t-test (2-tail) on RMF with the user-item bipartite network, RMF with user-tag bipartite network and WUDiff_RMF over the MAE and RMSE. The gains in accuracy are statistically significant with 95\% confidence level as displayed in Table 3. The results show that the improvements of WUDiff_RMF over the other two methods are statistically significant ($p < 0.01$) on four datasets.\par
In conclusion, the experimental results meet our expectation that utilizing the rating information and tagging information together to find the neighbour of users is better than only utilizing users' ratings or only utilizing users' taggings. And hence, by incorporating the neighborhood information from WUDiff into RMF, our method obtains significant promotion on prediction accuracy.\par

\subsubsection{Performance on Users with Different Number of Ratings and tags}
When users offer few ratings and few tags or even have no rating and tag, it is an critical challenge of the RSs to make an accurate prediction. In order to analyze the capabilities of our model thoroughly, we first group all the users based on the number of the observed ratings and the number of tag assignments in the training datasets, then make predictions and evaluate the prediction accuracies on different user groups. Additionally, to interpret the results more intuitively, we include the baseline method RMF for comparison since it does not incorporate any the neighborhood information.\par

\begin{figure}[H]
\subfigure[Delicious] {\includegraphics[width=0.5\textwidth]{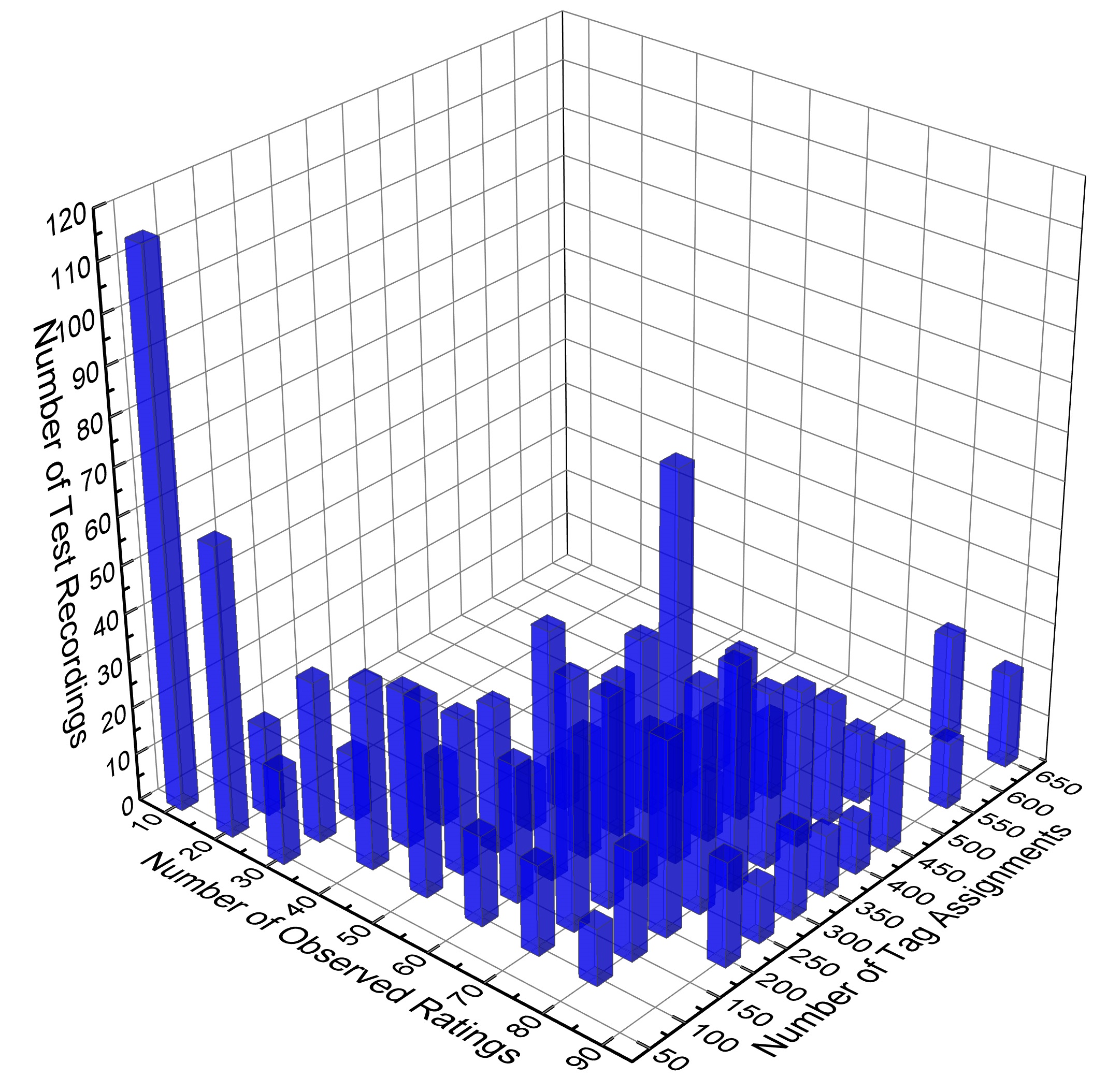}}
\subfigure[Last.fm] {\includegraphics[width=0.5\textwidth]{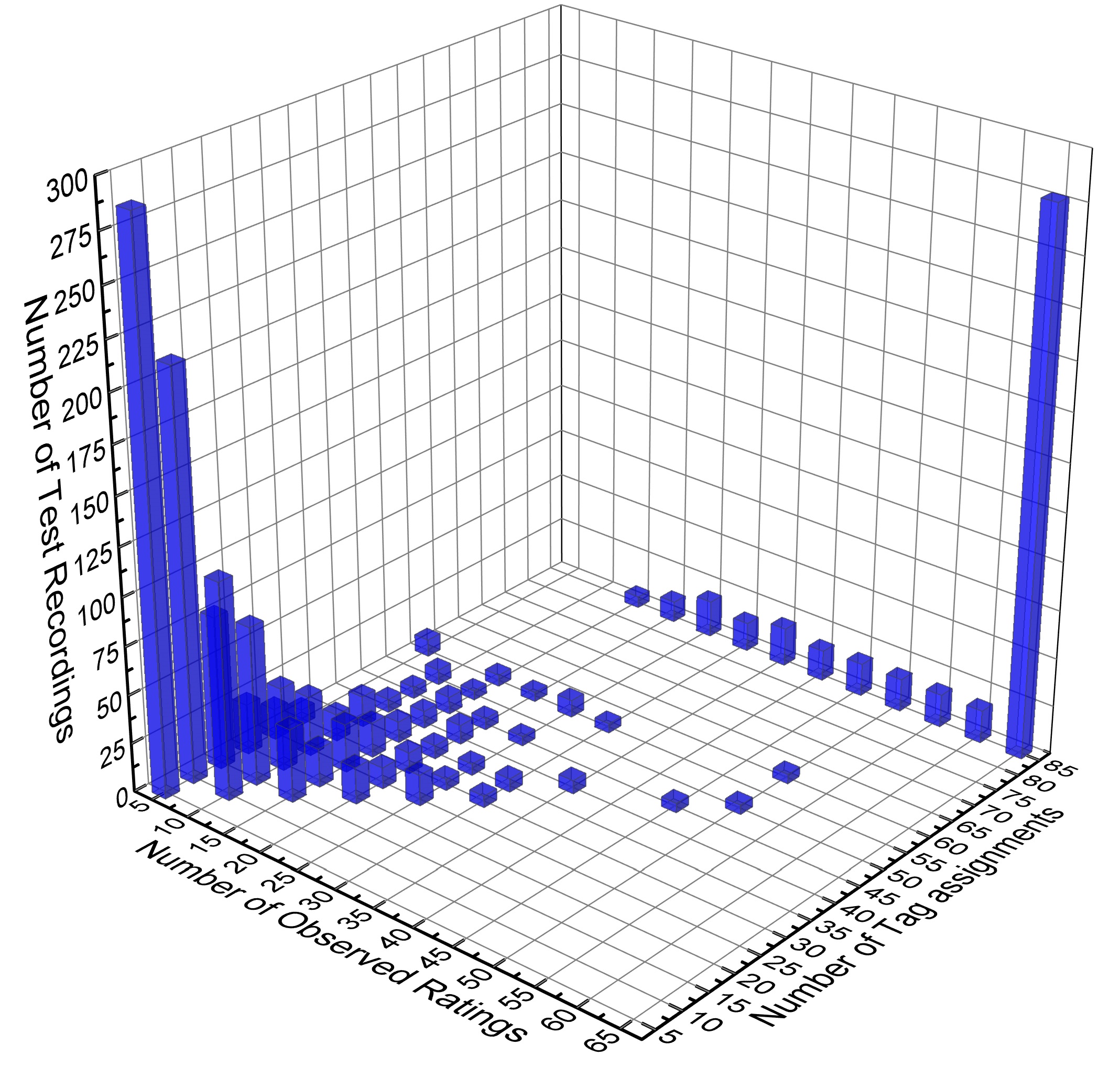}}
\subfigure[DBLP] {\includegraphics[width=0.5\textwidth]{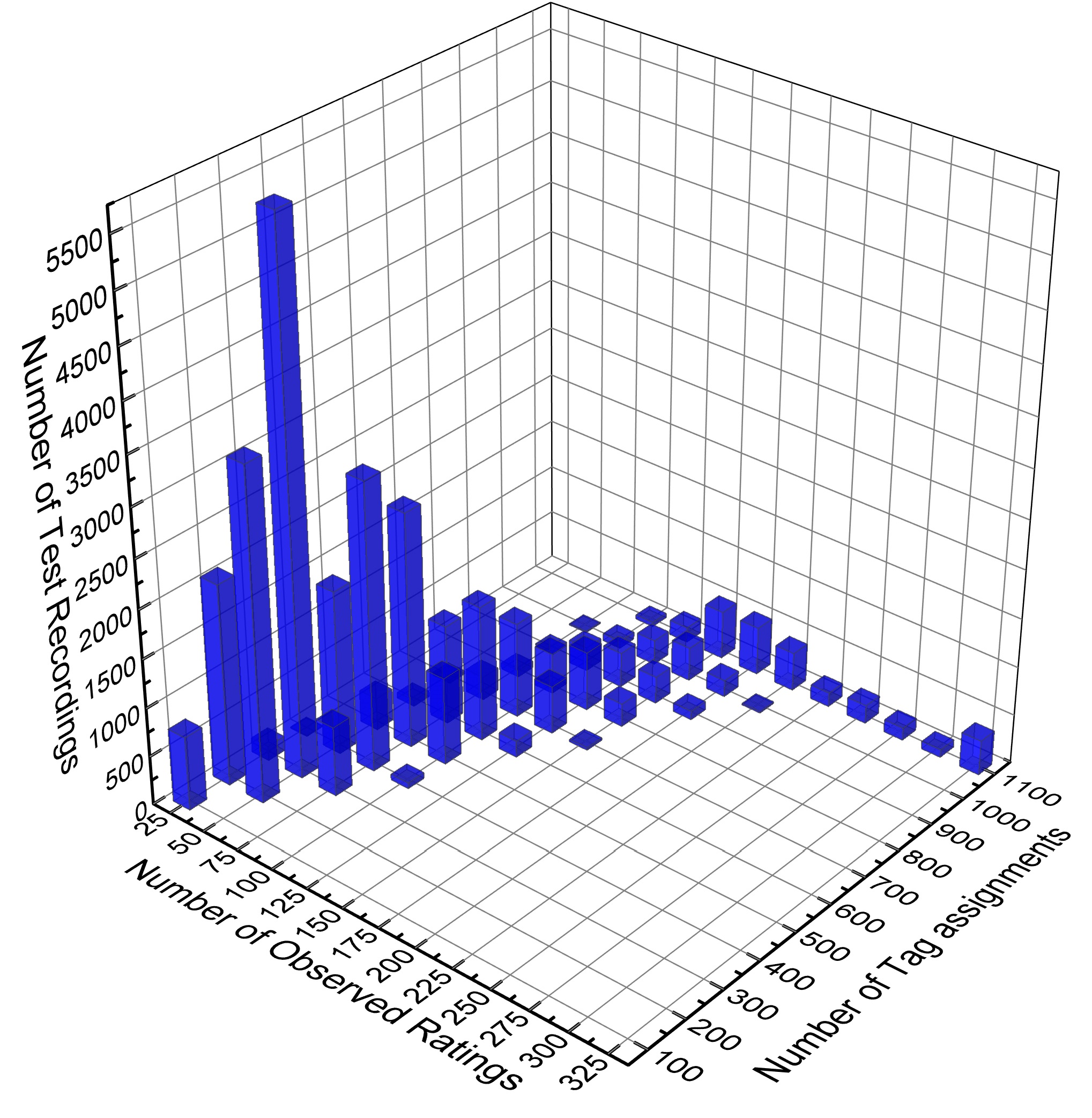}}
\subfigure[Movielens] {\includegraphics[width=0.5\textwidth]{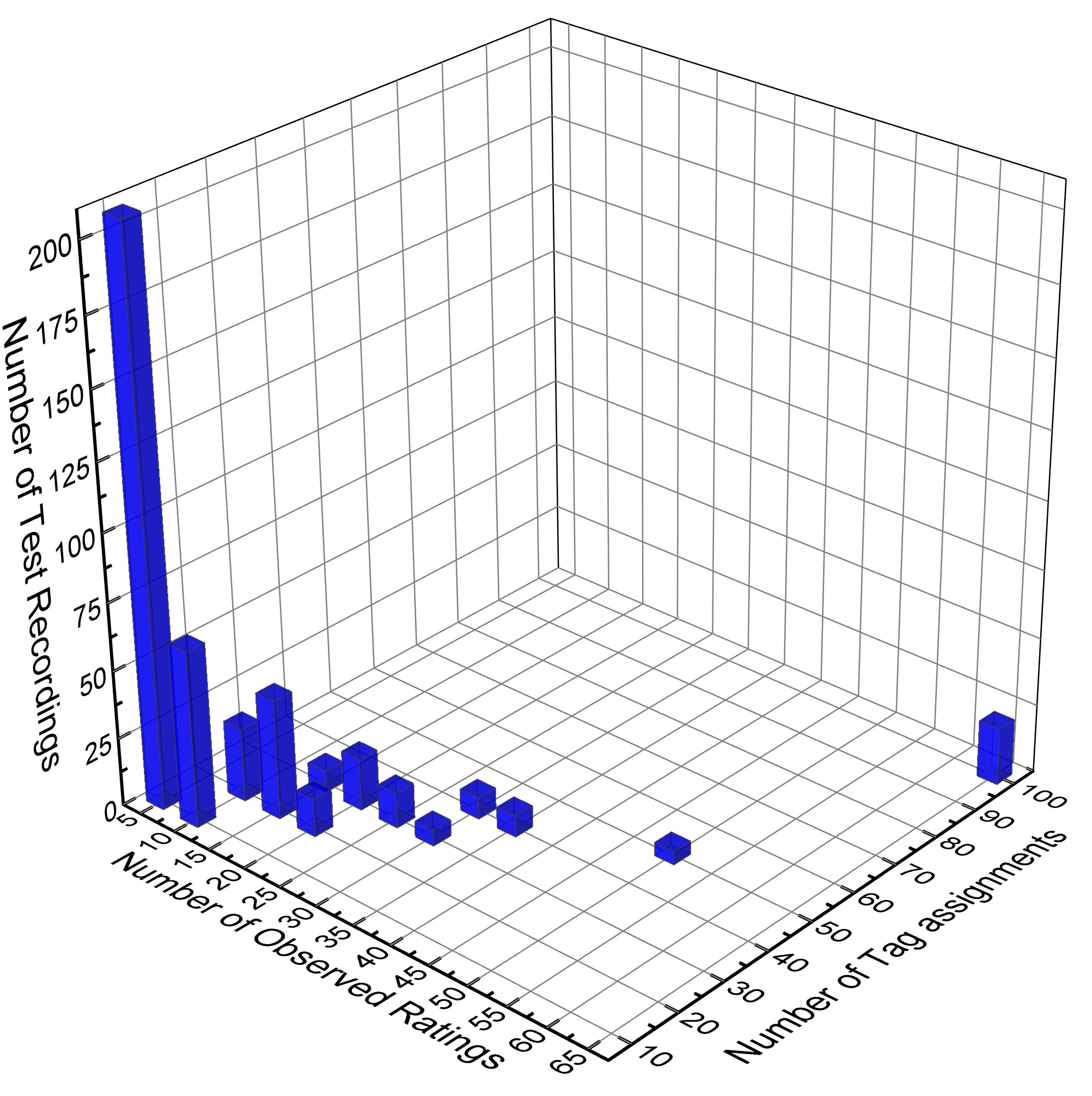}}
\caption{Distribution of Testing Data }
\label{fig:7}       
\end{figure}

Based on ratings and tag information, users are grouped into different classes. In MovieLens dataset, users are grouped into 13 classes: (5,10), (10,10), (10,20), (15,20), (15,30), (20,20), (20,30), (25,30), (30,30), (30,40), (35,40), (50,50) and ($>=65$,$>=100$). (5,10) denotes that the users have 0 to 5 ratings and 0 to 10 tags. There are 48, 55 and 46 categories in Delicious, Last.fm and DBLP datasets, separately. Fig.7 summarizes the user group distributions of the testing data in different datasets. For clarity, the results are illustrated by two-dimensional graph and three-dimensional histogram in Fig.8 and Fig.9. Since RMSE is more indicative than MAE, we only present the performance in RMSE when $f = 20$.\par

\begin{figure}[H]
\subfigure[RMSE(Delicious)] {\includegraphics[width=0.5\textwidth]{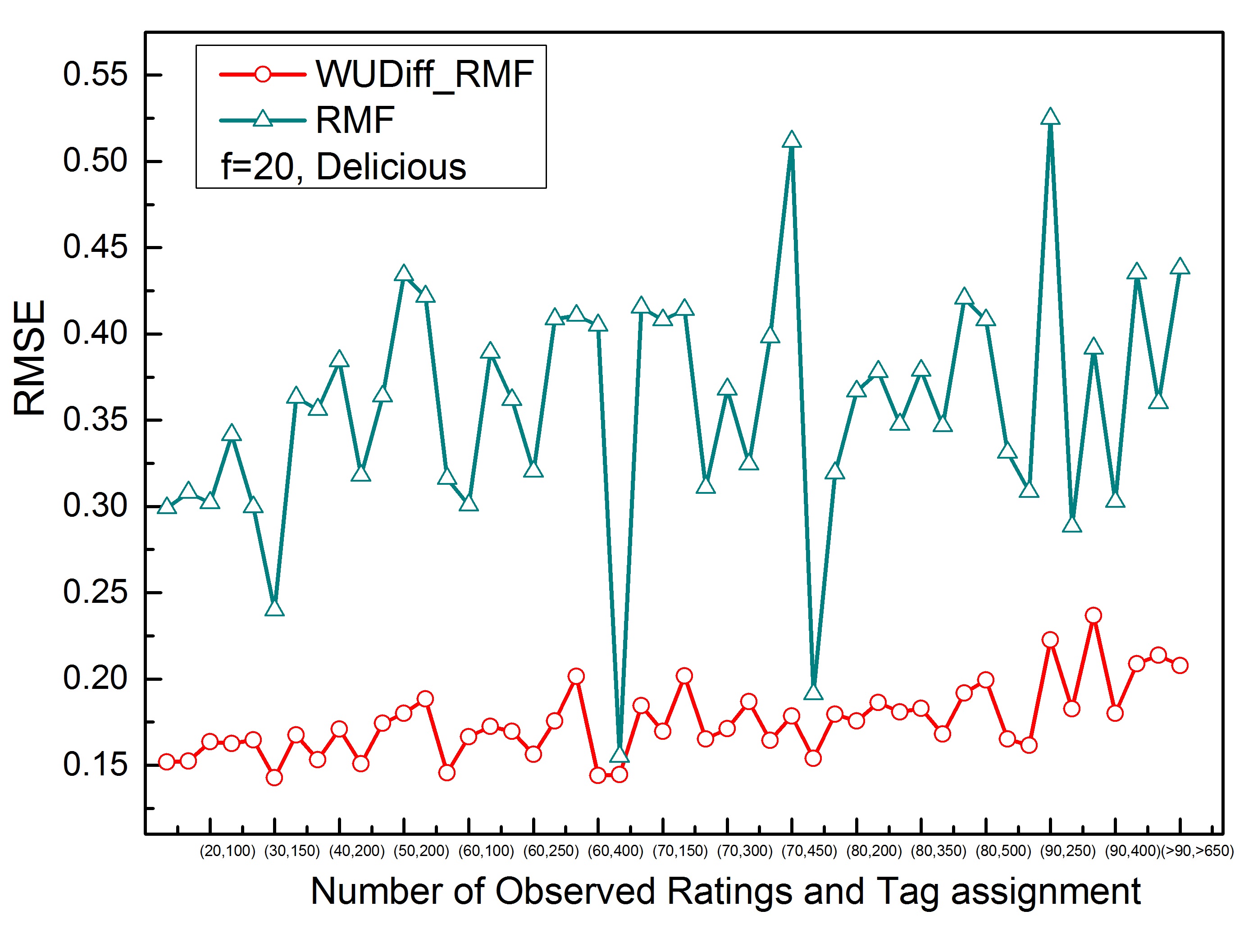}}
\subfigure[RMSE(Last.fm)] {\includegraphics[width=0.5\textwidth]{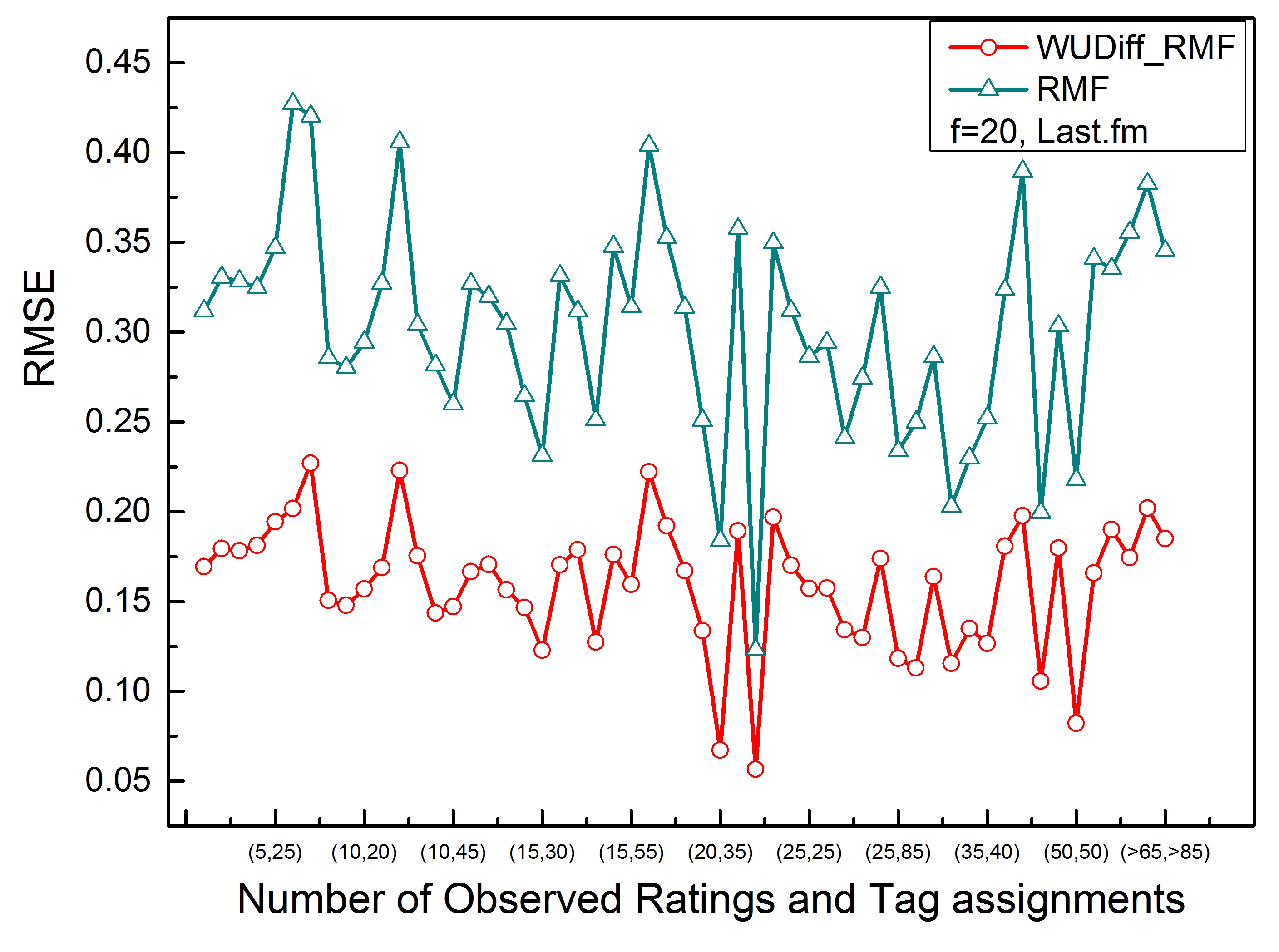}}
\subfigure[RMSE(DBLP)] {\includegraphics[width=0.5\textwidth]{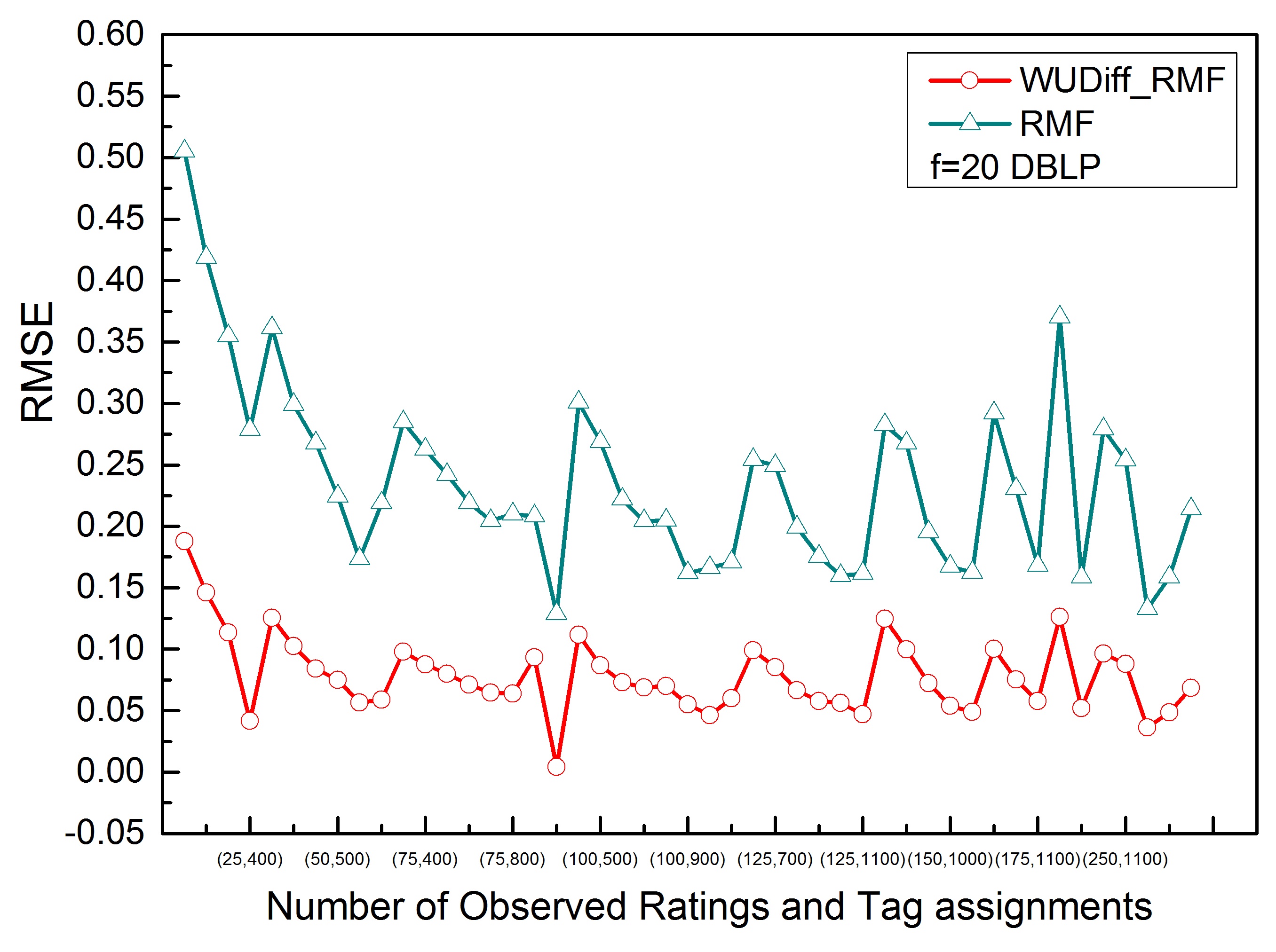}}
\subfigure[RMSE(Movielens)] {\includegraphics[width=0.5\textwidth]{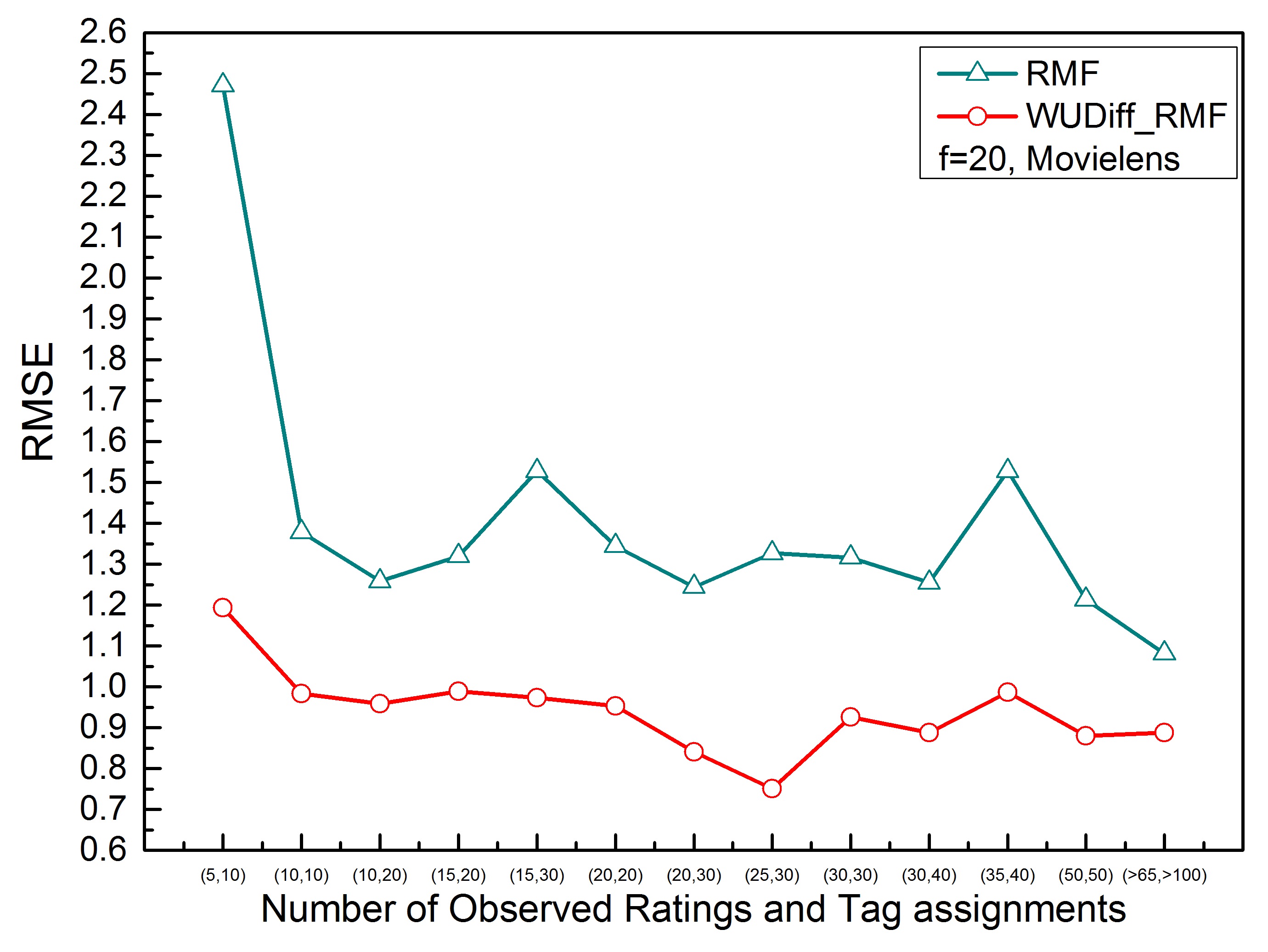}}
\caption{Performance comparison on different user groups (2-D graph) }
\label{fig:9}       
\end{figure}

\begin{figure}[H]
\subfigure[RMSE(Delicious)] {\includegraphics[width=0.5\textwidth,height=0.35\textheight]{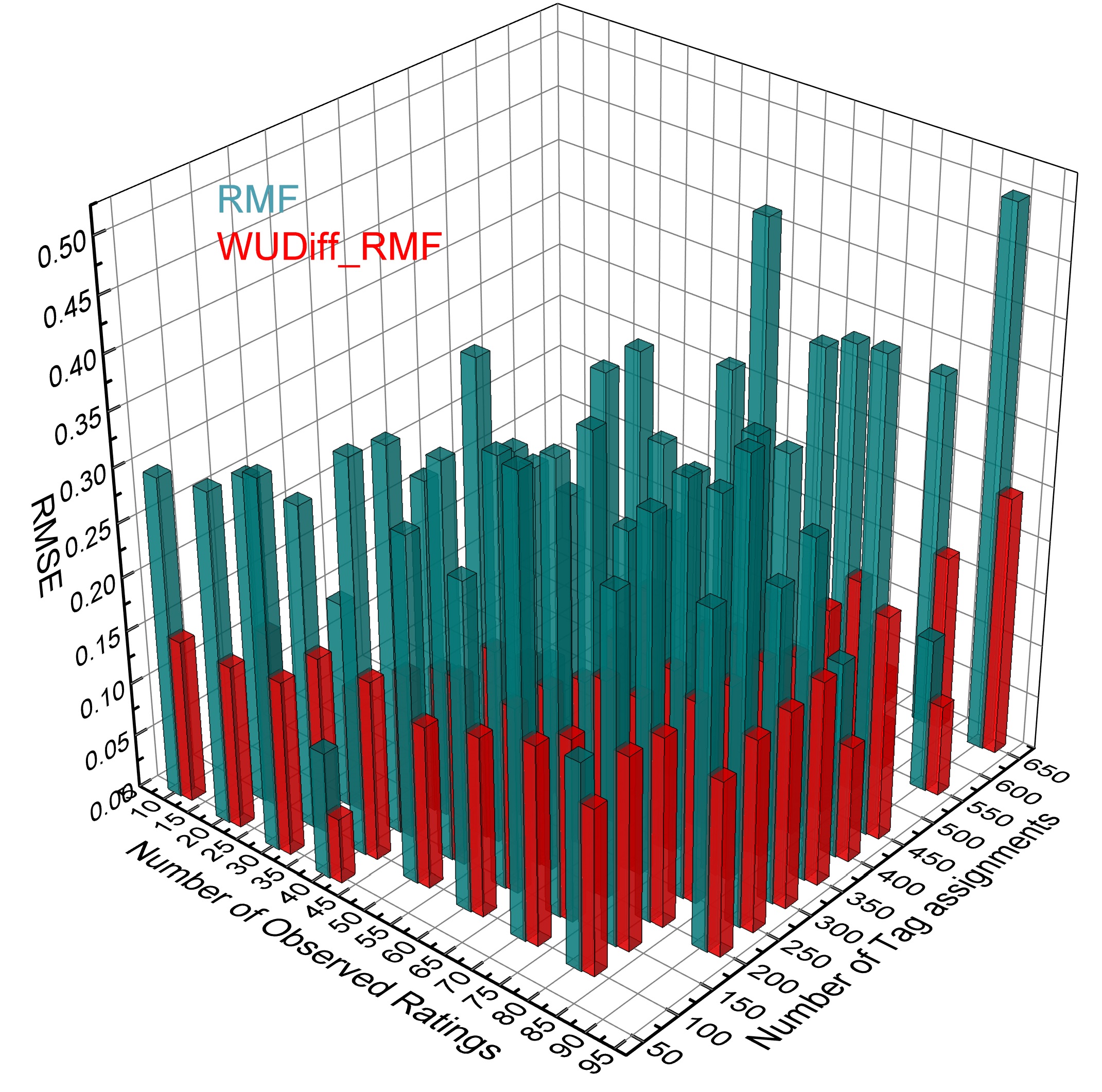}}
\subfigure[RMSE(Last.fm)] {\includegraphics[width=0.5\textwidth,height=0.35\textheight]{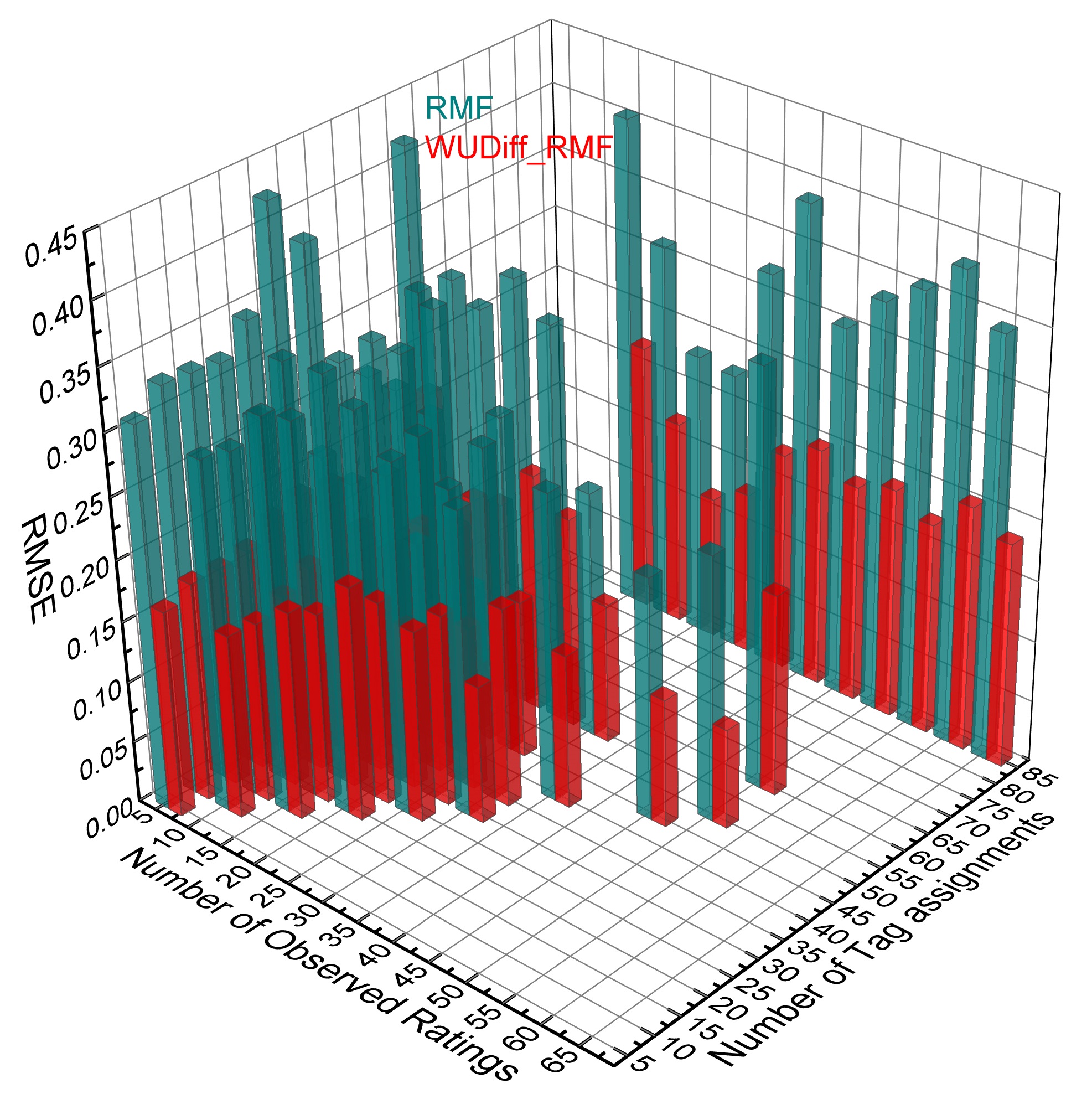}}
\subfigure[RMSE(DBLP)] {\includegraphics[width=0.5\textwidth,height=0.35\textheight]{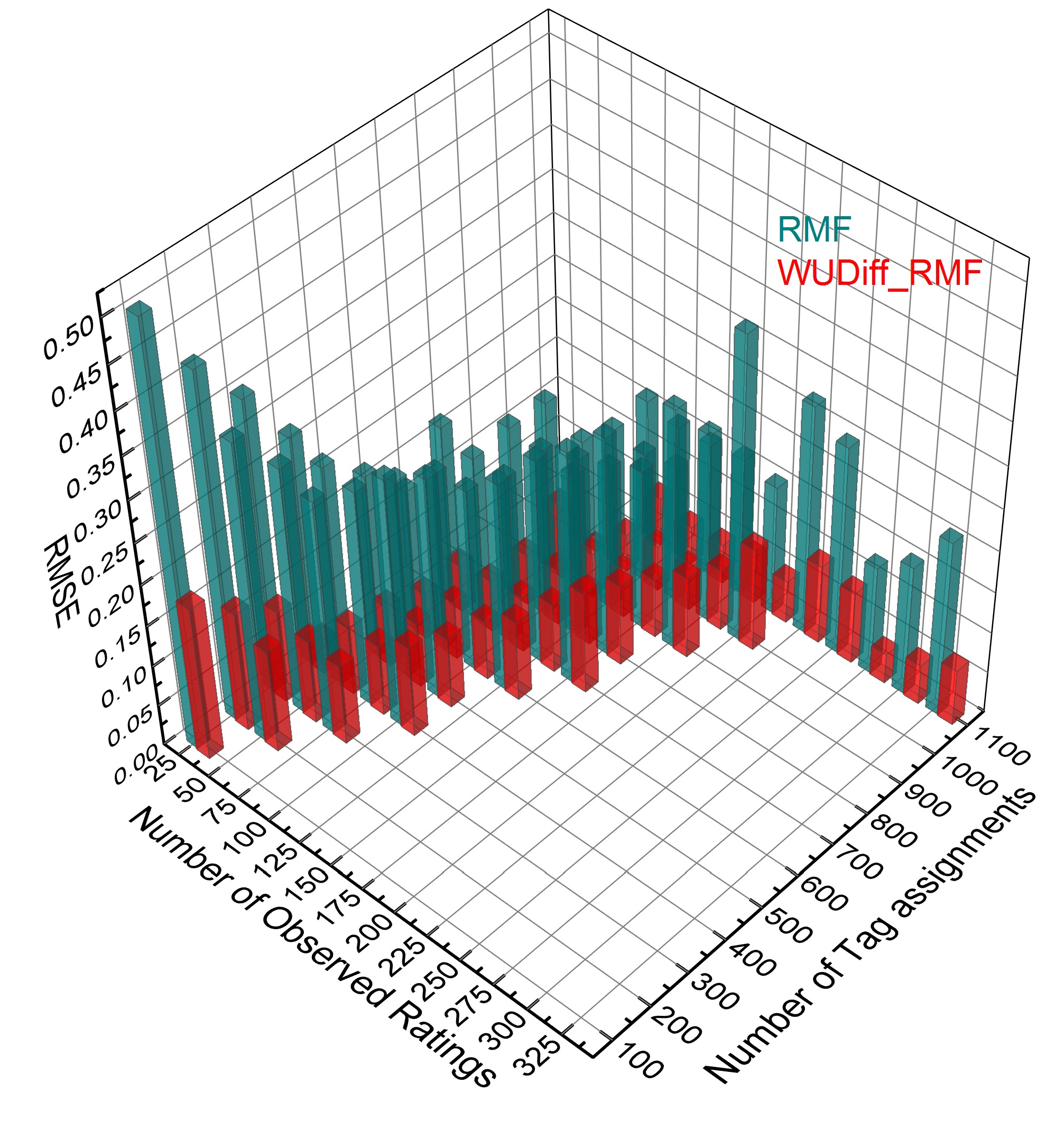}}
\subfigure[RMSE(Movielens)] {\includegraphics[width=0.5\textwidth,height=0.35\textheight]{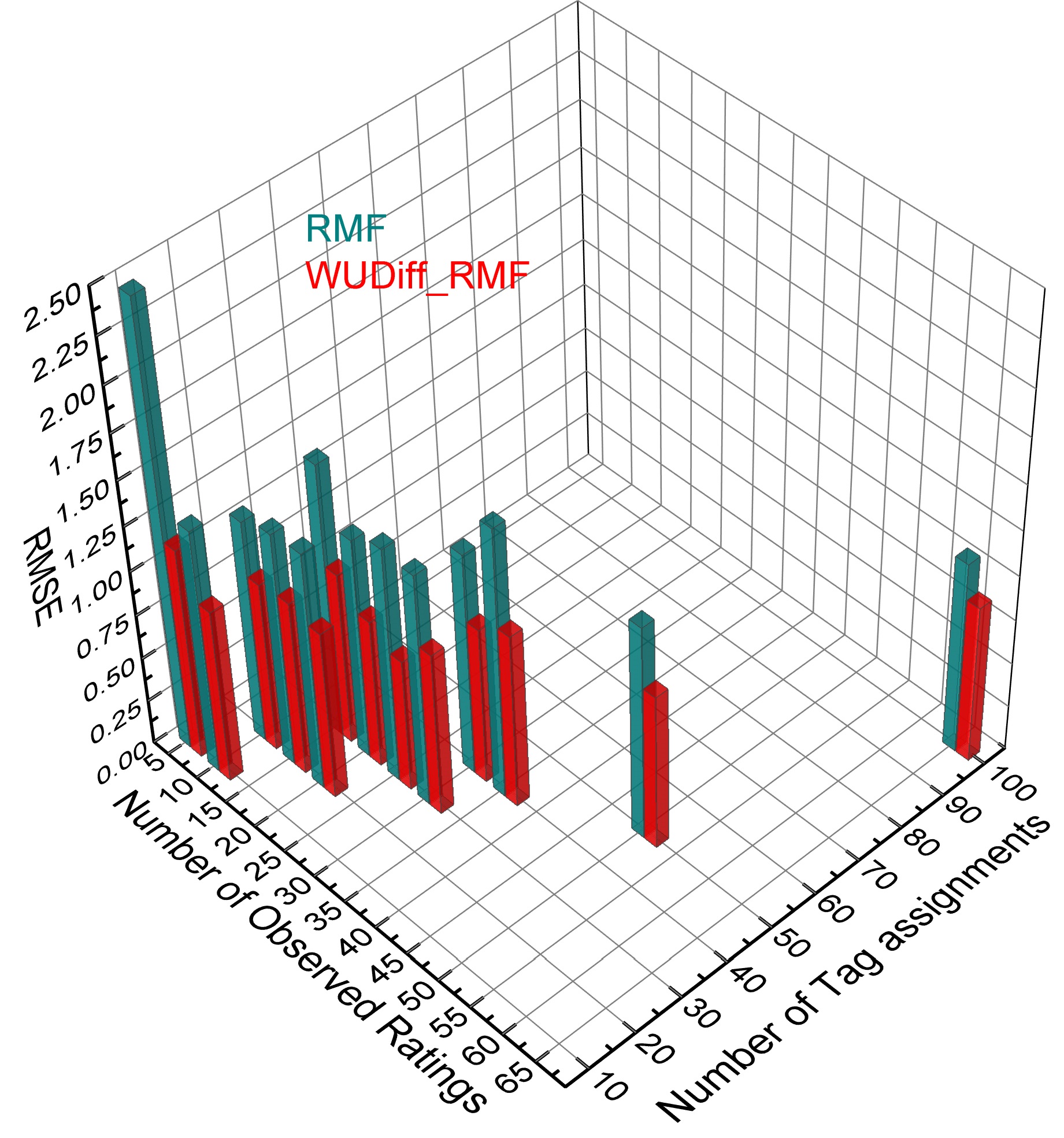}}
\caption{Performance comparison on different user groups (3-D histogram)}
\label{fig:8}       
\end{figure}

From these results, the performance of two compared methods changed with different user groups. The comparison demonstrates that our model can perform better than the pure RMF approach in all user groups. Even when the user groups have few ratings and few tags, especially the group with no more than 5 ratings or 10 tags, WUDiff_RMF also improves prediction accuracy. Actually, when little ratings are observed, pure RMF methods can not work well by only relying on the sparse rating matrix. Thus, fusing the implicit information of ¡°neighbors¡± based on the degree correlation of the tripartite network is a more efficient way to enhance accurate recommendations.

\section{Conclusion and Future Work}
\label{Conclusion and future work}
In this paper, based on the intuition that the users' ratings and tags have already reflected their interests, we present a hybrid recommendation model called WUDiff_RMF. A distinct feature of our model obtains the implicit information of similar users by utilizing the degree correlation of the user-item-tag tripartite network, and expands the RMF by integrating similar user regularization term. The experimental analysis of four datasets suggests that our proposed approach has better prediction accuracy over the existing seven methods, and the further analysis indicates it can alleviate the data sparsity, especially when active users have very few ratings and few tags.\par
There still exist several possible future extensions to our method. We will combine our model with other information of users' behaviour, such as time information, social trust network information and the semantic information of tags. Moreover, aiming at working well with larger datasets, it is necessary to adopt some parallel computing methods to speed up the calculation.

\section*{Acknowledgements}
The authors would like to thank Chenyi Zhang for his valuable datasets: DBLP and Movielens, and express their gratitude to Jing He for helpful suggestions.This work was partially supported by the following projects: National High-Technology Research and Development Program (¡°863¡± Program) of China(Grant No. 2013AA01A212); National Natural Science Foundation of China(GrantNo. 61272067, 61502180, 61370229); Natural Science Foundation of Guangdong Province of China (Grant No. S2012030006242); Natural Science Foundation of Guang dong Province of China(Grant No. 2014A030310238, 2016A030313441 ); Science and Technology Program of Guangzhou, China(Grant No.201508010067).




\end{document}